\relax
\documentclass[twoside]{article}
\usepackage[accepted]{aistats2019}

\usepackage{xr}

\usepackage{times}  %Required
\usepackage{helvet}  %Required
\usepackage{courier}  %Required
\usepackage{url}  %Required
\usepackage{graphicx}  %Required
\usepackage{amsmath}
\usepackage{amssymb}
\usepackage{pseudocode}
\usepackage{algorithm}
\usepackage{algorithmicx}
\usepackage{algpseudocode}
\usepackage{color}
\usepackage{tikz,pgf}

\usetikzlibrary{arrows,shapes.arrows,calc,shapes.geometric,shapes.multipart,  
decorations.pathmorphing,positioning,shapes.swigs}
\newenvironment{prf}{\noindent\textit{Proof:}\begin{mdseries}}{\end{mdseries}{\hfill\scriptsize$\Box$}} 

\newtheorem{thm}{Theorem}

\newtheorem{cor}{Corollary}
\newtheorem{prop}{Proposition}

\DeclareMathOperator{\doo}{do}

\DeclareMathOperator{\an}{an}

\DeclareMathOperator{\pa}{pa}

\DeclareMathOperator{\Pre}{Pre}
\DeclareMathOperator{\Pa}{Pa}
\DeclareMathOperator{\Ch}{Ch}

\DeclareMathOperator{\An}{An}
\DeclareMathOperator{\De}{De}

\def\ci{\perp\!\!\!\perp}

\begin{document}

% If your paper is accepted and the title of your paper is very long,
% the style will print as headings an error message. Use the following
% command to supply a shorter title of your paper so that it can be
% used as headings.
%
\runningtitle{A Potential Outcomes Calculus}

% If your paper is accepted and the number of authors is large, the
% style will print as headings an error message. Use the following
% command to supply a shorter version of the authors names so that
% they can be used as headings (for example, use only the surnames)
%
%\runningauthor{Surname 1, Surname 2, Surname 3, ...., Surname n}

\twocolumn[

\aistatstitle{A Potential Outcomes Calculus for Identifying\\ Conditional Path-Specific Effects}

\aistatsauthor{ Daniel Malinsky \And Ilya Shpitser \And Thomas Richardson }
%\aistatsauthor{ Author 1 \And Author 2 }

\aistatsaddress{ Johns Hopkins University \\ Department of Computer Science \\ Baltimore, MD USA \\ \texttt{malinsky@jhu.edu} \And Johns Hopkins University \\ Department of Computer Science \\ Baltimore, MD USA\\ \texttt{ilyas@cs.jhu.edu} \And University of Washington \\ Department of Statistics \\ Seattle, WA USA\\ \texttt{thomasr@uw.edu} } ]

\begin{abstract}
The do-calculus is a well-known deductive system for deriving connections between
interventional and observed distributions, and has been proven complete for a number of important identifiability problems in causal inference \cite{huang06do, pearl09causality, shpitser06id}. Nevertheless, as it is currently defined, the do-calculus is inapplicable to causal problems that involve complex nested counterfactuals which cannot be expressed in terms of the ``do'' operator.   Such problems include analyses of path-specific effects and dynamic treatment regimes.
%In addition, rule 3 of do-calculus is quite difficult to understand, in the sense that it involves three distinct interventional worlds.
In this paper we present the \emph{potential outcome calculus} (po-calculus), a natural generalization of do-calculus for arbitrary potential
outcomes. We thereby provide a bridge between identification approaches which have their origins in artificial intelligence and statistics, respectively. We use po-calculus to give a complete identification algorithm for
conditional path-specific effects with applications to problems in mediation analysis and algorithmic fairness.
\end{abstract}

\section{Introduction}

Pearl's do-calculus \cite{pearl94calculus, pearl95causal, pearl09causality} is an abstract set of rules for reasoning about interventions that has proven to be influential in settings, such as computer science and artificial intelligence, where graphical models are used to represent causal relationships.  In statistics and some social/biomedical sciences, the potential outcome framework \cite{neyman1923, rubin76inference} is more commonly used to express causal assumptions and reason about interventions. Richardson and Robins \cite{thomas13swig} have made an important contribution by unifying causal formalisms grounded in graphical causal models with the potential outcomes framework. In this paper we build on those connections, presenting a calculus for reasoning about interventions in the potential outcomes notation that is equivalent to Pearl's do-calculus for standard interventions, but allows generalizations to nested
causal quantities pertinent to evaluating (e.g.)\ dynamic treatment regimes or path-specific interventions (for which the ``do'' notation is insufficiently expressive).  We show how the new calculus can be applied to problems in mediation analysis, specifically the identification of conditional path-specific causal effects. We introduce a procedure which is complete for expressing such quantities as functions of the observed data distribution, i.e., an algorithm which will produce an identifying expression for a conditional path-specific effect if and only if the effect is identifiable. %{\color{red}maybe need example early.}

\begin{figure*}
	\begin{center}
		\begin{tikzpicture}[>=stealth, node distance=1.2cm]
		\tikzstyle{format} = [draw, very thick, circle, minimum size=5.0mm,
		inner sep=0pt]
		\tikzstyle{unode} = [draw, very thick, circle, minimum size=1.0mm,
		inner sep=0pt]
		\tikzstyle{square} = [draw, very thick, rectangle, minimum size=4mm]
		
		\begin{scope}[xshift=0.0cm]
		\path[->, very thick]
		node[] (dummy) {}
		node[format, below of=dummy, yshift=0.4cm] (c) {$C$}
		node[format, right of=c] (w) {$A$}
		node[format, right of=w] (m) {$M$}
		node[format, right of=m] (y) {$Y$}
		
		node[format, gray, above of=w, yshift=-0.5cm] (h1) {$H_1$}
		node[format, gray, below of=w, xshift=0.6cm, yshift=+0.5cm] (h2) {$H_2$}
		
		(c) edge[blue] (w)
		(w) edge[blue] (m)
		(m) edge[blue] (y)
		
		(h1) edge[red] (c)
		(h1) edge[red] (m)
		
		(h2) edge[red] (c)
		(h2) edge[red] (y)
		
		(w) edge[blue, bend left] (y)
		
		node[below of=w, yshift=-0.3cm, xshift=0.6cm] (l) {$(a)$}
		;
		\end{scope}
		
		\begin{scope}[xshift=5.0cm]
		\path[->, very thick]
		node[] (dummy) {}
		node[format, below of=dummy, yshift=0.4cm] (c) {$C$}
		node[format, right of=c] (w) {$A$}
		node[format, right of=w] (m) {$M$}
		node[format, right of=m] (y) {$Y$}
		
		%			node[format, gray, above of=w, yshift=-0.5cm] (h1) {$H_1$}
		node[format, gray, below of=w, xshift=0.6cm, yshift=+0.5cm] (h2) {$H_2$}
		
		(c) edge[blue] (w)
		(w) edge[blue] (m)
		(m) edge[blue] (y)
		
		%			(h1) edge[red] (c)
		%			(h1) edge[red] (m)
		
		(c) edge[blue, bend left] (m)
		
		(h2) edge[red] (c)
		(h2) edge[red] (y)
		
		(w) edge[blue, bend left] (y)

		node[below of=w, yshift=-0.3cm, xshift=0.6cm] (l) {$(b)$}
		;
		\end{scope}
		
		\end{tikzpicture}
	\end{center}
	\caption{(a) A hidden variable causal DAG where $p(Y(a,M(a')))$ is identified, but $p(Y(a,M(a')) \mid C)$ is not identified.
		(b) A seemingly similar hidden variable causal DAG where both $p(Y(a,M(a')))$ and $p(Y(a,M(a')) \mid C)$ are identified.
	}
	\label{fig:example}
\end{figure*}
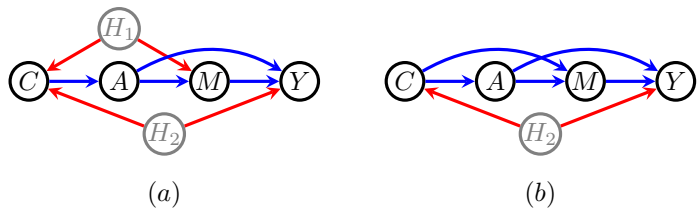

%Pearl introduced the ``do'' operator notation to describe interventional distributions, e.g., $p(Y \mid \text{do}(a))$ for the distribution of $Y$ when $A$ is assigned to $a$ by an intervention. We use the potential outcome notation where $p(Y(a)) \equiv p(Y \mid \text{do}(a))$. 
Conditional path-specific effects are quantified via conditional distributions over potential outcomes, where treatment variables are assigned to possibly distinct values for different causal pathways.
 %that behave as if one or more treatment variables were set to different values for different causal pathways from treatments to outcomes.  
In mediation analysis,
functions of such distributions are used to isolate the effect of a drug, therapy, or other treatment assignment along a specific pathway
in a specific subpopulation, defined by pre-treatment variables (such as age or gender) or post-treatment variables (such as adverse reactions to the treatment). Importantly, there are settings where the marginal path-specific effect is identified but the conditional path-specific effect is not identified; we later discuss one simple example shown in Fig.~\ref{fig:example}. 

%A conditional path-specific quantity like $p(Y(a,M(a')) \mid W(a,M(a')))$ can then be used to narrow mediation analysis to a particular subpopulation, e.g., patients who have specific values for their baseline covariates or would have post-treatment diagnostic variables within some range. This can be particularly important in precision medicine.

Another context in which conditional path-specific effects may be of interest is in the study of algorithmic fairness. Recent papers \cite{nabi18policy, nabi18fair, Zhang17causal} have proposed to combat disparities perpetuated by some automated decision-making systems by identifying, estimating, and constraining unfair causal influences that propagate along certain pathways, e.g., the direct effect of gender on hiring outcomes or the indirect effect of race on criminal justice outcomes via geographical factors. It may also be desirable to constrain such path-specific effects for certain subpopulations, which requires identifying conditional path-specific effects.

We begin by introducing potential outcomes, causal models, graphs, and some relevant results. Then we review the do-calculus, propose our potential outcome calculus, demonstrate they are equivalent, and give some simple derivations to establish the soundness of the rules in the language of potential outcomes. Finally, we introduce a formalism for expressing path-specific effects (PSEs) and a complete identification procedure for conditional PSEs.  %We defer the proofs of some claims to the Supplement in the interests of space.

%Pearl introduced the ``do'' operator notation to describe interventional distributions, e.g., $p(Y \mid \text{do}(a))$ for the distribution of $Y$ when $A$ is assigned to $a$ by an intervention. We use the potential outcome notation where $p(Y(a)) \equiv p(Y \mid \text{do}(a))$. A benefit of using the potential outcome notation for the purposes of this paper is that we can express causal quantities which involve complex, nested sequences of assignments, and conditional distributions like XXX which for which the ``do'' notation is insufficient.

\section{Potential Outcomes, the Do Operator and Causal Models}
Fix a set of indices $K \equiv \{ 1, \ldots, k \}$ under a total ordering $\prec$.  For each random variable $V_i$, $i \in K$, define a state space ${\mathfrak X}_i$, and the sets $\Pre_{i} \equiv \{ 1, \ldots, i-1 \}$.  Given $A \subseteq K$, we will denote subsets of random variables indexed by $A$ with $V_A$ and elements $v_A$ of ${\mathfrak X}_A$ by $a$ (lowercase letters).

We assume the existence of all one-step-ahead \emph{potential outcome} random variables (a.k.a.\ counterfactuals) of the form $V_i(\pa_i) \equiv V_i(v_{\Pa_{i}})$, where $\Pa_{i}$ is a fixed subset of $\Pre_{i}$, and $\pa_i \equiv v_{\Pa_{i}}$ is any element in ${\mathfrak X}_{\Pa_{i}}$.  The variable $V_i({\pa_{i}})$ denotes the value of $V_i$ had the set of \emph{direct causes of $V_i$}, or $V_{\Pa_{i}}$, been set, possibly contrary to fact, to values ${\pa_{i}}$.  The existence of a total ordering $\prec$ on indices, and the fact that $\Pa_{i} \subseteq \Pre_{i}$ precludes the existence of cyclic causation. (That is, we consider causal models that are \emph{recursive}.)
$V_i({\pa_{i}})$ may be conceptualized as the output of a \emph{structural equation}
$f_i : {\mathfrak X}_{\Pa_{i} \cup \{ \epsilon_i \}} \mapsto {\mathfrak X}_i$, a function representing a causal mechanism that maps values of $\Pa_{i}$, as well as the value of an exogenous disturbance variable $\epsilon_i$, to values of $V_i$.
We define causal models as sets of densities over the set of random variables
\begin{align*}
\mathbb{V} \equiv \{ V_i({\pa_{i}}) \mid i \in \{ 1, \ldots, k \}, {\pa_{i}} \in {\mathfrak X}_{\Pa_{i}} \}.
\end{align*}
For simplicity of presentation, we assume ${\mathfrak X}_i$ is always finite, and thus ignore the measure theoretic complications that arise with defining densities over sets of random variables above in the case where some state spaces on $\Pa_{i}$ are infinite.\footnote{The set of $p({\mathbb V})$ for a particular set of $\Pa_i$ and an ordering $\prec$ was called the \emph{finest causally interpretable structured tree graph (FCISTG)} in \cite{robins86new}.}

Given a set of one-step-ahead potential outcomes $\mathbb{V}$, for any $A \subseteq K$ and $i \in K$ we define the potential outcome $V_i(a)$, the response of $V_i$ had variables in $V_A$ been set to $a$, by the definition known as \emph{recursive substitution}:
\begin{align}
V_i(a) \equiv V_i({a \cap \pa_i}, \{ V_j(a) \mid j \in \Pa_i \setminus A \}).
\label{eqn:rec-sub}
\end{align}
In words, this states that $V_i(a)$ is the potential outcome where variables $\Pa_i$ in $A$ are set to their corresponding values in $a$, and all elements of $\Pa_i$ not in $A$ are set to whatever values their recursively defined counterfactual versions would have had had $A$ been set to $a$.  Equivalently, $V_i(a)$ is the random variable induced by a modified set of structural equations: specifically the set of functions $f_j$ 
%$f_j : {\mathfrak X}_{\Pa_j \cup \{ \epsilon_j \}} \to {\mathfrak X}_{j}$ 
for all $V_j \in A$ are replaced by constant functions $f^*_j$ that set $V_j$ to the corresponding value in $a$.

We denote by ${\mathbb V}^*$ the set of all variables derived by (\ref{eqn:rec-sub}) from ${\mathbb V}$, together with ${\mathbb V}$.
In addition, for notational conciseness, we will use index sets to denote sets of potential outcomes themselves.  That is, for $Y \subseteq K$, $A \subseteq K$, we will denote the set $\{ V_i(a) \mid i \in Y \}$ by $Y(a)$.  Note that we allow $Y$ and $A$ to intersect.  Thus, we allow sets of potential outcomes of the form $A(a)$, which denote the sets $\{ V_i(a) \mid i \in A \}$, where each $V_i(a)$ is defined using
(\ref{eqn:rec-sub}) above.  In particular, if $A = \{ V_i \}$ (a singleton), $V_i(v_i)$ is defined in our notation to be the random variable $V_i$, not the constant $v_i$.

%Pearl used the ``do'' operator notation to describe interventional distributions, e.g.,\ $p(Y \mid \doo(a))$ for the distribution of $Y$ when $A$ is assigned to value $a$ by an intervention. Using the potential outcome notation we identify $p(Y(a)) \equiv p(Y \mid \text{do}(a))$ where $Y(a)$ is the value of $Y$ under assignment $A=a$, described in more detail below. As noted in \cite{thomas13swig}, there is no way to express quantities like $p(Y(a) \mid A = a')$ in the ``do'' notation. In mediation analysis, functions of quantities like $p(Y(a,M(a')))$ (with a \emph{mediating} variable $M$) may be of interest, since these can be used to isolate the effect of a drug, therapy, or other treatment assignment along a specific pathway. Examples include the effect of smoking on lung cancer via nicotine exposure rather than smoke inhalation, or the effect of HIV treatment on viral failure via biochemical pathways rather than differential adherence.

In cases where $Y$ and $A$ do not intersect, the distribution $p(Y(a))$ has been denoted by Pearl as $p(Y \mid \doo(a))$ \cite{pearl09causality}.  This formulation places emphasis on the intervention operator
$\doo(a)$, which replaces structural equations by constants.
%While this view has a number of advantages, it also has disadvantages, as we discuss below.

%We call all distributions on $\mathbb{V}$ and $\mathbb{V}^*$ obtained from $\mathbb{V}$ via (\ref{eqn:rec-sub}) \emph{recursive structural equation models (RSEMs)}.  These models are very general models on potential outcomes that preclude cyclic causation.

Recursive substitution provides a link between observed variables and potential outcomes. In particular, it implies the \emph{consistency property}:\footnote{Some readers may be more familiar with the simpler formulation where $a = \emptyset$, so ``$B = b\text{ implies } V_i(b) = V_i$.'' Our reasons for allowing multiple intervention sets will become clear in what follows.}
%in RSEMs:
for any disjoint $A,B \subseteq K$, $i \in K \setminus (A \cup B)$, $a \in {\mathfrak X}_A$, $b \in {\mathfrak X}_B$,
\begin{align}
B(a) = b\text{ implies } V_i(a,b) = V_i(a).
\label{eqn:consist}
\end{align}
%See \cite{thomas13swig}.
\begin{prop}[consistency]
	Given ${\mathbb V}^*$ derived from ${\mathbb V}$ via (\ref{eqn:rec-sub}), then (\ref{eqn:consist}) holds.
\label{prop:consistency}
\end{prop}
\begin{prf}
	By (\ref{eqn:rec-sub}), $V_i(a)$ and $V_i(a,b)$ are defined as
	\vspace{-2mm}
	\begin{align*}
%	V_i(a) \equiv \hspace{6.53cm}\\ %\!\!\equiv\!\!
	V_i({a_{\Pa_i}}, \!\{ V_j(a) | j \!\in\! \Pa_i \!\setminus\! ( A \cup B ) \}\!, \!\{ V_j(a) \!\!=\!\! b_j | j \!\in\! B \!\cap\! \Pa_i \}\!)\\
	%V_i(a,b) \equiv
	\text{and }V_i({a_{\Pa_i}}, \!\{ V_j(a,b) | j \!\in\! \Pa_i \!\setminus\! ( A \cup B ) \}, \! {b_{\Pa_i}}), \hspace{4mm}
	\end{align*}
	respectively.
	The conclusion follows immediately.
\end{prf}

%In addition,
(\ref{eqn:rec-sub}) implies that every $V_i(a)$ is can be written as a function of a unique minimally causally relevant
subset of $a$.
\begin{prop}[causal irrelevance]
Given ${\mathbb V}^*$ derived from ${\mathbb V}$ via (\ref{eqn:rec-sub}), let $V_i(a) \in \mathbb{V}^*$,
and let $A^*$ be the maximal subset of $A$ such that for every $A_j \in A^*$,
there exists a sequence $V_{w_1}, \ldots, V_{w_m}$ that does not intersect
$A$, where $A_j \in \Pa_{w_1}$, $V_{w_i} \in \Pa_{w_{i+1}}$, for $i = 1, \ldots m-1$, and $V_{w_m} \in \Pa_i$.
Then $V_i(a) = V_i(a^*)$.
\label{prop:causal-relevance}
\end{prop}
\begin{prf}
Follows by definition of $A^*$ and (\ref{eqn:rec-sub}).
\end{prf}

A \emph{functional causal model} (a.k.a.\ a non-parametric structural equation model with independent errors, NPSEM-IE) asserts that the sets of variables
\begin{align}
\left\{
\{ V_i({\pa_i}) \mid {\pa_i} \in {\mathfrak X}_{\Pa_i} \} \mid i \in \{ 1, \ldots, k \}
\right\}
\label{eqn:mwm}
\end{align}
are mutually independent.  Phrased in terms of structural equations, the functional causal model states that the joint distribution of the disturbance terms factorizes into a product of marginals:
%\begin{align*}
$p(\epsilon_1, \ldots, \epsilon_k) = \prod_{i=1}^k p(\epsilon_i).$
%\end{align*}

Alternative causal models, which make fewer assumptions than the functional model but are sufficient for all inferences we aim to make in this paper, are discussed in \cite{thomas13swig, shpitser15hierarchy}.  We focus on the functional causal model here, since it is simpler to describe and the original setting of Pearl's do-calculus. We discuss how our results apply to a weaker causal model
\cite{thomas13swig} in the Supplement.
% Section \ref{sec:weaker}.
%{\color{red}Revisit later.}

\section{Graphical Models}

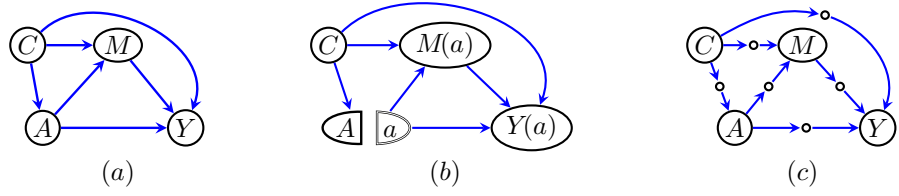
\begin{figure*}
	\begin{center}
		\begin{tikzpicture}[>=stealth, node distance=1.2cm]
		\tikzstyle{format} = [draw, thick, circle, minimum size=4.0mm,
		inner sep=1pt]
		\tikzstyle{unode} = [draw, thick, circle, minimum size=1.0mm,
		inner sep=0pt,outer sep=0.9pt]
		\tikzstyle{square} = [draw, very thick, rectangle, minimum size=4mm]
		\begin{scope}
		\path[->,  line width=0.9pt]
		node[format] (c) {$C$}
		node[format, shape=ellipse, right of=c, xshift=0.0cm] (m) {$M$}			
		node[format, below of=c, xshift=0.2cm,yshift=0.1cm] (a) {$A$}
		node[format, xshift=0.7cm, right of=a] (y) {$Y$}
		
		(c) edge[blue] (a)
		(a) edge[blue] (y)
		(a) edge[blue] (m)
		(m) edge[blue] (y)
		(c) edge[blue] (m)
		(c) edge[blue, out=45, in=70, looseness=1.1] (y)
		%(c) edge[blue] (y)
		
		node[below of=m, yshift=-0.5cm, xshift=0.0cm] (l) {$(a)$}
		;
		\end{scope}
		
		\begin{scope}[xshift=4.0cm]
		\path[->, line width=0.9pt]
		node[format] (c) {$C$}
		node[format, shape=ellipse,right of=c, xshift=0.35cm] (m) {$M\!(a)$};
		\begin{scope}
			\tikzset{line width=0.9pt, inner sep=1.8pt, swig vsplit={gap=6pt, inner line width right=0.3pt}}	
				\node[below of=c, xshift=0.5cm, yshift=0.1cm, name=a, shape=swig vsplit]{
        					\nodepart{left}{$A$}
        					\nodepart{right}{$a$} };
		\end{scope}
		\path[->, thick]
		node[format, shape=ellipse, right of=a, xshift=1.0cm] (y) {$Y\!(a)$}
		
		(c) edge[blue] (a)
		(a) edge[blue] (y)
		(a) edge[blue] (m)
		(m) edge[blue] (y)
		(c) edge[blue] (m)
		(c) edge[blue, out=45, in=70, looseness=1.1] (y)
		
		node[below of=m, yshift=-0.5cm, xshift=0.0cm] (l) {$(b)$}
		;
		\end{scope}
		
		\begin{scope}[xshift=9.0cm]
		\path[->, thick]
		node[format] (c) {$C$}
		node[format, shape=ellipse, right of=c, xshift=0.1cm] (m) {$M$}			
		node[format, below of=c, xshift=0.4cm,yshift=0.1cm] (a) {$A$}
		node[format, xshift=0.7cm, right of=a] (y) {$Y$}

%		node[format] (c) {$C$}
%		node[format, below of=c, yshift=0.1cm] (m) {$M$}			
%		node[format, below left of=m, yshift=0.2cm] (a) {$A$}
%		node[format, below right of=m, yshift=0.2cm] (y) {$Y$}
		
		node[unode] (ca) at ($(c)!0.5!(a)$) {} %calc package
		node[unode] (cm) at ($(c)!0.5!(m)$) {}
		
		node[unode] (am) at ($(a)!0.5!(m)$) {}
		node[unode] (ay) at ($(a)!0.5!(y)$) {}
		
		node[unode] (my) at ($(m)!0.5!(y)$) {}
		
		node[unode, above of=m, yshift=-0.8cm, xshift=0.3cm] (cy){}
		
		(c) edge[blue] (ca)
		(ca) edge[blue] (a)
		
		(c) edge[blue] (cm)
		(cm) edge[blue] (m)
		
		(a) edge[blue] (am)
		(am) edge[blue] (m)
		
		(a) edge[blue] (ay)
		(ay) edge[blue] (y)
		
		(m) edge[blue] (my)
		(my) edge[blue] (y)
		
		(c) edge[blue,out=30,in=170] (cy)
		(cy) edge[blue,out=-20,in=70] (y)
		
		node[below of=m, yshift=-0.5cm, xshift=0.0cm] (l) {$(c)$}
		;
		\end{scope}
		
		%\begin{scope}[xshift=12.0cm]
		%\path[->, very thick]
		%node[] (dummy) {}
		%node[format, below of=dummy, yshift=0.4cm] (c) {$C$}
		%node[format, right of=c] (w) {$W$}
		%node[format, right of=w] (m) {$M$}
		%node[format, right of=m] (y) {$Y$}
		
		%node[format, gray, above of=w, yshift=-0.5cm] (h1) {$H_1$}
		%node[format, gray, below of=w, xshift=0.6cm, yshift=+0.5cm] (h2) {$H_2$}
		
		%(c) edge[blue] (w)
		%(w) edge[blue] (m)
		%(m) edge[blue] (y)
		
		%(h1) edge[red] (c)
		%(h1) edge[red] (m)
		
		%(h2) edge[red] (c)
		%(h2) edge[red] (y)
		
		%node[below of=w, yshift=-0.3cm, xshift=0.6cm] (l) {$(d)$}
		%;
		
		%\end{scope}
		
		\end{tikzpicture}
	\end{center}
	\caption{(a) A simple causal DAG $\cal G$, with a treatment $A$, an outcome $Y$, a vector $C$ of baseline variables, and a mediator $M$.
		(b) A SWIG ${\cal G}(a)$ derived from (a) corresponding to the world where $A$ is intervened on to value $a$.
		(c) An extended graph ${\cal G}^e$ derived from (a).
		%(d) A hidden variable DAG illustrating a fact implied by the causal model not inferrable from do-calculus alone: that $p(Y(w) \mid M(w))$ is not a function of $w$.
	}
	\label{fig:triangle}
\end{figure*}

Much conceptual clarity may be gained by viewing causal models as graphs.  We will consider graphs with either directed edges only ($\to$), or mixed graphs with both directed and bidirected ($\leftrightarrow$) edges. Vertices correspond to random variables, and we simplify notation by using $V_i$ to refer to both the graph vertex and corresponding random variable. In all cases we will require the absence of directed cycles, meaning that whenever the graph contains a path of the form $V_i \to \cdots \to V_j$, the edge
$V_j \to V_i$ cannot exist.  Directed graphs with this property are called directed acyclic graphs (DAGs), and mixed graphs with this property are called acyclic directed mixed graphs (ADMGs).  We will refer to graphs by ${\cal G}(V)$, where $V$ is the set of random variables indexed by $\{ 1, \ldots, k \}$.  We will use the following standard definitions for sets of vertices in a graph:
\begin{align*}
\Pa_i^{\cal G} &\equiv \{ V_j \mid V_j \to V_i \text{ in }{\cal G} \} & \text{ (parents of $V_i$)}\\
\An_i^{\cal G} &\equiv \{ V_j \mid V_j \to \cdots \to V_i \text{ in }{\cal G} \} & \text{ (ancestors of $V_i$)}\\
\De_i^{\cal G} &\equiv \{ V_j \mid V_j \gets \cdots \gets V_i \text{ in }{\cal G} \} & \text{ (descendants of $V_i$)}
%\Sib_i &\equiv \{ V_j \mid V_j \leftrightarrow V_i \text{ in }{\cal G} \} & \text{ (siblings of $V_i$)}\\
%\Dis_i &\equiv \{ V_j \mid V_j \leftrightarrow \ldots \leftrightarrow V_i \text{ in }{\cal G} \} & \text{ (the district of $V_i$)}.
\end{align*}
By convention, we assume $V_i \in \An_i^{\cal G}$ and $V_i \in \De_i^{\cal G}$.
%As we define them neither $\An_i^{\cal G}$ nor $\De_i^{\cal G}$ include $V_i$.
We will generally drop the superscript ${\cal G}$ if the relevant graph is obvious and sometimes write ${\cal G}$ in place of ${\cal G}(V)$ when the vertex set is clear.
%For any $A \subseteq V$, ${\cal G}_A$ is the induced subgraph of ${\cal G}$, containing only vertices in $A$, and edges in ${\cal G}(V)$ between $A$.  Given a graph ${\cal G}(V)$, define the set of districts in ${\cal G}(V)$ by ${\cal D}({\cal G}(V))$.  Note that districts always partition $V$.
Given a DAG ${\cal G}(V)$, a statistical DAG model (a.k.a.\ a Bayesian network) associated with ${\cal G}(V)$ is a set of distributions that are Markov relative to ${\cal G}(V)$, i.e., the set of distributions that can be written as the following product of conditional densities:
\begin{align}
p(V) = \prod_{i = 1}^k p(V_i \mid \Pa_i).
\label{eqn:d}
\end{align}
Given $p(V)$ that is Markov relative to a DAG $\mathcal{G}(V)$, conditional independence relations (written: $(Y \ci Z \mid X)$, where $X,Y,Z$ are disjoint subsets of the index set $K$) satisfied by $p(V)$ can be derived using the well-known d-separation criterion \cite{pearl88probabilistic}, which we reproduce in the Supplement. We write $(Y \ci_d Z \mid X)_{{\cal G}(V)}$ when $Y$ is d-separated from $Z$ given $X$ in ${\cal G}(V)$.
If $p(V)$ is Markov relative to ${\cal G}(V)$, then the following \emph{global Markov property} holds: for any disjoint $X,Y,Z$
\[
(Y \ci_d Z \mid X)_{{\cal G}(V)} \Rightarrow (Y \ci Z \mid X) \text{ in } p(V).
\]
Functional causal models may also be associated with a DAG ${\cal G}$ by identifying $\Pa_i$ with the graphical parents of $V_i$ in ${\cal G}(V)$. Given a functional causal model for DAG $\mathcal{G}$, the joint distribution for any $V(a)$ derived from $\mathbb{V}$ using (\ref{eqn:rec-sub}) is identified via the following formula:
\begin{align}
p(V(a)) = %\prod_{i \not\in A}
\prod_{i=1}^K p(V_i \mid \Pa_i \setminus A, a \cap \pa_i),
\label{eqn:g}
\end{align}
provided $\left( \prod_{a_j \in a} p(a_j \mid \Pa_j \setminus A, a \cap \pa_j) \right) > 0$.
See \cite{thomas13swig} for a simple proof.
The modified factorization (\ref{eqn:g}) is known as the \emph{extended g-formula}
\cite{thomas13swig, robins04effects}.  Note that (\ref{eqn:g}) has a term for every $V_i \in V$, just like (\ref{eqn:d}).

The formula (\ref{eqn:g}) resembles (\ref{eqn:d}) and in fact may be viewed as a factorization of $p(V(a))$ with respect to a certain graph derived from ${\cal G}$.  Such graphs, called Single World Intervention Graphs (SWIGs), were introduced in \cite{thomas13swig}.  SWIGs are graphical representations of potential outcome densities that help unify the graphical and potential outcome formalisms. Given a set $A$ of variables which are assigned to values $a$, a SWIG ${\cal G}(a)$ is constructed
from ${\cal G}(V)$ by splitting all vertices in $A$ into a random half and a fixed half, with the random half inheriting all  edges with an incoming arrowhead and the fixed half inheriting all outgoing directed edges.  Then, all random vertices $V_i$ are re-labelled as $V_i(a)$ or equivalently (due to Proposition \ref{prop:causal-relevance}) as $V_i(a \cap \an^*_i)$, where $\an^*_i$ consists of values of the ancestors of $V_i$ in the split graph.
%{\color{red} values of the ancestors of $V_i$ along all directed paths without two consecutive edges of the form $\to A_i \to$, where $A_i \in A$. [This sentence still doesn't make sense to me. First I presume you mean $A_j$ since $i$ is the index of $V_i$. More importantly ``ancestors along all paths'' is vague here. I believe ``ancestors of $V_i$ in the split graph'' is more obviously what is written in the SWIG paper.]}
In \cite{thomas13swig}, unsplit vertices were drawn as circles, and split nodes as half circles, with fixed nodes denoted by a lowercase.  
Fixed nodes are enclosed by a double line.
For an example of a SWIG representing the joint density $p(Y(a), M(a), C(a), A(a)) = p(Y(a),  M(a), C, A)$, see Fig.~\ref{fig:triangle} (b).
Because of the resemblance of (\ref{eqn:g}) to a DAG factorization, we say that $p(V(a))$ is Markov relative to a SWIG ${\cal G}(a)$ if
$p(V(a))$ may be written as (\ref{eqn:g}).  

A SWIG ${\cal G}(a)$ is a DAG with a vertex set $\{V(a),a\}$, and may be viewed as a \emph{conditional} graph, with vertices in $V(a)$ corresponding to random variables, and vertices in $a$ corresponding to variables fixed to a value. We extend the notion of d-separation to allow fixed vertices.  Specifically, we allow d-separation statements of the form $(Y(a),a' \ci_d Z(a) \mid X(a))_{{\cal G}(a)}$, for disjoint random subsets $Y(a),Z(a),X(a)$ of $V(a)$ and $a'$ a subset of $a$. Note that a possibly d-connecting path may only contain random nodes as non-endpoint vertices (as in \cite{thomas13swig} where fixed nodes are always blocked). Our extension here consists only in allowing fixed vertices to also appear as one endpoint in a d-separation statement.
%{\color{red}Note that in a SWIG, we implicitly assume $a \setminus a'$ is conditioned on in this statement. [Explain? Or remove?]}
Just as (\ref{eqn:d}) implied the global Markov property for a DAG, the modified factorization (\ref{eqn:g}) implies a global Markov property for a SWIG.
%That is, d-separation between random nodes implies a corresponding conditional independence between potential outcome variables and d-separation of fixed nodes from random nodes implies that corresponding conditional distributions are \emph{not} functions of the separated (fixed) set.

\begin{prop}[SWIG global Markov property] \label{prop:extended-cdag-markov}
If $p(V(a))$ is Markov relative to ${\cal G}(a)$, then for any disjoint subsets $Y(a),Z(a),X(a)$ of $V(a)$ and a subset $a'$ of $a$,
if $(Y(a), a' \ci_d Z(a) \mid X(a))_{{\cal G}(a)}$ then, for some $f(\cdot)$,
%there exists $f()$ such that
\begin{align*}
p(Z(a) | Y(a), X(a)) &= p(Z(a) | X(a)) = f(Z,X,a \setminus a'). %\\
%&=p(B(a \setminus a') \mid C(a \setminus a')) \text{ if $C(a)$ is empty}.
\end{align*}
%for some $f(\cdot)$.
\end{prop}
\begin{prf}
The first equality is due to Theorem 12 in \cite{thomas13swig}, the second follows from Theorem 19 in \cite{richardson17nested}. \end{prf}

Note that $f(Z,X,a \setminus a')$ is not necessarily equal to $p(Z(a \setminus a') \mid X(a \setminus a'))$.
%We discuss the implications of this fact for do-calculus in the Supplement.
%For example, in the SWIG ${\cal G}(w)$ derived from Fig.~\ref{fig:triangle} (d), $(Y(w) \ci w \mid M(w))$, and indeed $p(Y(w) \mid M(w)) = f(Y,M)$.  However, $p(Y(w) \mid M(w))$ is not equal to $p(Y \mid M)$ under this causal model. 
%We defer some relevant results about SWIGs to the Supplement.

The SWIG global Markov property implies the following intuitive result (proved in the Supplement)
relating independence statements in $p(V(a))$ for various sets $A$.
%, that factorize as (\ref{eqn:g})
Specifically, the result is that interventions ``always help'' when it comes to conditional independence.

\begin{prop}[intervention monotonicity]
	For any disjoint subsets $Y(a),Z(a),X(a)$ of $V(a)$ and a subset $a'$ of $a$,
	if $(Y(a), a' \ci Z(a) \mid X(a))_{{\cal G}(a)}$ then for any $A'' \supseteq A$,
	$(Y(a''), a' \ci Z(a'') \mid X(a''))_{{\cal G}(a'')}$.
	\label{prop:do-always-helps}
\end{prop}

\subsection*{Graphical Models With Hidden Variables}

We also consider causal models where some variables are unmeasured (a.k.a.\ ``latent'' or ``hidden'' variables). Given a DAG ${\cal G}(V \cup H)$, define a \emph{latent projection} mixed graph ${\cal G}(V)$ as follows.  $V$ is the vertex set of ${\cal G}(V)$, and for any $V_i,V_j \in V$ there is an edge $V_i \to V_j$ if there exists a directed path from $V_i$ to $V_j$ in ${\cal G}(V \cup H)$, with all intermediate nodes on the path in $H$; there is an edge $V_i \leftrightarrow V_j$ if there exists a path from $V_i$ to $V_j$
of the form $V_i \gets \cdots \to V_j$, where every intermediate node on the path is in $H$ and no consecutive edges on the path are of the form $\to H_k \gets$ for $H_k \in H$.  The latent projection ${\cal G}(V)$ obtained from a DAG ${\cal G}(V \cup H)$ is always an ADMG.  
%In addition, most identification theory in causal inference is shared by any two causal models associated with DAGs ${\cal G}_1(V \cup H_1)$, ${\cal G}_2(V \cup H_2)$ that yield the same latent projection ${\cal G}_1(V) = {\cal G}_2(V)$.  In this sense, the latent projection ADMG defines a natural class of causal models.  
Our results in this paper apply to ADMGs, and indeed this is the intended setting for Pearl's do-calculus (he used the terminology ``semi-Markovian models'').
%If $p(V \cup H)$ factorizes as (\ref{eqn:d}) with respect to
%${\cal G}(V \cup H)$, ${\cal G}(V)$ obeys the nested Markov factorization with respect to ${\cal G}(V)$, with details described in the appendix.

%The notation for latent projections is meant to be analogous to notation for marginal distributions.  As we will see, ${\cal G}(V)$ is a ``graphical marginal'' of ${\cal G}(V \cup H)$.
%\section*{Global Markov Properties}

The definition of d-separation naturally generalizes to ADMGs with minor modification for bidirected edges; the resulting criterion is called
m-separation \cite{richardson03markov}.
%, the same is true if the arrowhead-based definition of d-separation is appropriately amended to allow for bidirected edges;
%the resulting criterion is called m-separation }; m-separation reduces to d-separation if $\mathcal{G}$ is a DAG.
%We reproduce the definitions in the Appendix.
We write\\ $(Y \ci_m Z \mid X)_{{\cal G}(V)}$ if $Y$ is m-separated from $Z$ given $X$ in ADMG $\mathcal{G}(V)$. In the following we sometimes drop the $d$ or $m$ subscripts and just write $\ci$, where the relevant criterion is implicit.

%do we want to say anything about $V(a)$ living in an ADMG SWIG?  It _is_ well defined wrt a DAG ${\cal G}(V \cup H)$, so maybe not?

Given an ADMG ${\cal G}(V)$, we define a SWIG ${\cal G}(V)(a)$ by the analogous node splitting construction as for DAGs.
Specifically, each node is split into a random half and  a fixed half, with random halves inheriting all incoming directed and bidirected edges, and fixed halves inheriting all outgoing directed edges.  Alternatively given a SWIG ${\cal G}(a)$ derived from a DAG ${\cal G}(V \cup H)$, we define the latent projection operation in the natural way, yielding the SWIG ${\cal G}(a)(V)$ with random vertices $V$, fixed vertices $a$, and directed
edges from $a_i \in a$ or $V_i \in V$ to $V_j \in V$ if there is a directed path from the corresponding vertices in ${\cal G}(a)$
%the original graph 
with all intermediate vertices in $H$, and bidirected edges from $V_i \in V$ to $V_j \in V$ if there exists a path from $V_i$ to $V_j$
of the form $V_i \gets \ldots \to V_j$, where every intermediate node on the path is in $H$ and no consecutive edges on the path are of the form $\to H_k \gets$ for $H_k \in H$.  These operations commute, and we can derive independence statements via m-separation on $G(V)(a)$, as we prove in the Supplement.

%Note also that \cite{thomas13swig} focus their attention SWIGs constructed from DAGs (incl.\ hidden variable DAGs) whereas here we explicitly allow for SWIGs constructed from ADMGs.  The construction for ADMGs is exactly the same node splitting construction as for DAGs.
%{\color{red}describe how?}

\section{Do-Calculus and Potential Outcomes Calculus}

Pearl formulated the do-calculus originally as follows:
	\begin{align*}
	1\text{ : }&  p(y \mid z,w,\doo(x)) = p(y \mid w,\doo(x))\\
	 & \hspace{40mm} \text{ if } (Y \ci Z \mid W,X)_{{\cal G}_{\overline{X}}} \\
	2\text{ : }& p(y \mid z,w,\doo(x)) = p(y \mid w,\doo(z),\doo(x))\\
	 & \hspace{40mm} \text{ if } (Y \ci Z \mid W,X)_{{\cal G}_{\overline{X},\underline{Z}}} \\
	3\text{ : }& p(y \mid w,\doo(z),\doo(x)) = p(y \mid w, \doo(x))\\
	 & \hspace{40mm} \text{ if } (Y \ci Z \mid W,X)_{{\cal G}_{\overline{X},\overline{Z(W)}}}
	\end{align*}
where ${\cal G}_{\overline{X}}$ denotes the graph obtained from ${\cal G}$ by removing all edges with arrowheads into $X$, ${\cal G}_{\underline{Z}}$ denotes the graph obtained from ${\cal G}$ by removing all directed edges out of $Z$, and $Z(W) \equiv Z \setminus \An_{{\cal G}_{\overline{X}}}(W)$.

Here we present the do-calculus entirely in terms of potential outcomes (the ``potential outcomes calculus'' or ``po-calculus'' for short). The conditions are phrased in terms of conditional independencies implied by SWIGs, e.g., ${\cal G}(x)$ for the SWIG where $X$ is assigned value $x$. We restate the rules as follows:
	\begin{align*}
	1\text{ : }& p(Y(x) \mid Z(x), W(x)) = p(Y(x) \mid W(x))\\
	 & \hspace{20mm} \text{ if } (Y(x) \ci Z(x) \mid W(x))_{{\cal G}(x)} \\
	 %& \text{causal global Markov property on random variables.}\\
	2\text{ : }& p(Y(x,z) \mid W(x,z)) = p(Y(x) \mid W(x), Z(x) = z) \\
	 & \hspace{20mm} \text{ if } (Y(x,z) \ci Z(x,z) \mid W(x,z))_{{\cal G}(x,z)}\\
	 %& \text{generalized conditional ignorability.}\\
	3\text{ : }& p(Y(x,z) \mid W(x,z)) = p(Y(x) \mid W(x))\\
	 & \hspace{20mm} \text{ if } (Y(x,z_1), W(x,z_1) \ci z_1)_{{\cal G}(x,z_1)} \text{ and } \\
	& \hspace{20mm}	(Y(x,z_1) \ci Z_2(x,z_1) \mid W(x,z_1))_{{\cal G}(x,z_1)}\\
	& \hspace{20mm} \text{ where } Z_1 = Z \setminus \An_{{\cal G}(x)}(W),\\
	& \hspace{20mm} Z_2 = Z \cap \An_{{\cal G}(x)}(W)
	% 3^*\text{: }& p(Y(x,z)) = p(Y(x)) &\text{ if }&(Y(x,z) \ci_d z)_{{\cal G}(x,z)}
	\end{align*}
Recall that random variables in a SWIG ${\cal G}(x)$ are labelled $V_i(x)$ or equivalently as $V_i(x \cap \an_i^*)$, where
$\an^*_i$ consists of values of the ancestors of $V_i$ in the split graph.
%where $Z = Z_1 \dot{\cup} Z_2$, $Z_1 = Z \setminus \An_{{\cal G}(x)}(W)$, and $Z_2 = Z cap An_{G(x)}(w)$
%$Z_2 = Z \setminus Z_1$ in Rule $3$.
We can view Rule $1$ as the fragment of the SWIG global Markov property that pertains to random variables in $V(a)$. Rule $2$ may be called ``generalized conditional ignorability'' because it is a general version of the standard ignorability assumption used in causal inference settings, where $(Y(a) \ci A \mid C)$, or equivalently $(Y(a) \ci A(a) \mid C(a))$, enables identification of (e.g.) the average treatment effect by adjusting for $C$.
Note that Rule $3$ does not have a simple interpretation, as it involves an equality of interventional distributions in two distinct ``worlds,'' given an independence condition in a third. However, below we suggest an alternative, simpler rule which may be used without loss of generality, and is more intuitive. First, we state some basic results.

\begin{prop}
	Rule $1$ of po-calculus holds if and only if Rule $1$ of do-calculus holds.
\end{prop}
\begin{prf}
	Follows from the definition of ${\cal G}(x)$ and ${\cal G}_{\overline{X}}$, and the definition of m-separation. % in ${\cal G}(x)$.
\end{prf}

\begin{prop}
	Rule $2$ of po-calculus holds if and only if Rule $2$ of do-calculus holds.
\end{prop}
\begin{prf}
	Follows from the definition of ${\cal G}(x,z)$ and ${\cal G}_{\overline{X},\underline{Z}}$, and the definition of m-separation in ${\cal G}(x,z)$.
\end{prf}

\begin{prop}
	Rule $3$ of po-calculus holds if and only if Rule $3$ of do-calculus holds.
\end{prop}
\begin{prf}
	Since path separation criteria on graphs quantify over elements in vertex sets, and since $Z$ is a disjoint union of $Z_1$ ($Z(W)$ in Pearl's terminology) and $Z_2$,
the precondition in Rule $3$ of do-calculus may be written as two preconditions: $(Y \ci Z_1 \mid W,X)_{{\cal G}_{\overline{X},\overline{Z_1}}}$ and $(Y \ci Z_2 \mid W,X)_{{\cal G}_{\overline{X},\overline{Z_1}}}$.

By definition of $Z_1$, it contains only non-ancestors of $W$ in ${\cal G}_{\overline{X}}$ (and therefore also in ${\cal G}_{\overline{X},\overline{Z_1}}$, which is an edge subgraph of ${\cal G}_{\overline{X}}$).
Since $Z_1$ only has adjacent outgoing directed arrows in ${\cal G}_{\overline{X},\overline{Z_1}}$, all elements of $W$ are marginally m-separated from $Z_1$ in ${\cal G}_{\overline{X},\overline{Z_1}}$.  Thus, $(W(x,z_1) \ci z_1)_{{\cal G}(x,z_1)}$ by the definition of ${\cal G}(x,z_1)$.
%Thus, $(Y \ci Z_1 \mid W,X)_{{\cal G}_{\overline{X},\overline{Z_1}}}$ implies that $W$ is either an ancestor of $Z_1$ or disconnected from $Z_1$ in ${\cal G}$. In either case $(W(x,z_1) \ci z_1)_{{\cal G}(x,z_1)}$ by the definition of ${\cal G}(x,z_1)$.
	Furthermore, no element of $Z_1$ can be an ancestor of $Y$ in ${\cal G}_{\overline{X},\overline{Z_1}}$. To see this, suppose an element $Z_i$ of $Z_1$ were an ancestor of $Y$. Then since $(Y \ci Z_1 \mid W,X)_{{\cal G}_{\overline{X},\overline{Z_1}}}$, the directed path from $Z_i$ must be blocked by $W$ and $X$. $W$ cannot be on this directed path because it is non-descendant of $Z_1$, and $X$ cannot be on the path because ${\cal G}_{\overline{X},\overline{Z_1}}$ has no directed edges into $X$. So we conclude that $Z_i$ is not an ancestor of $Y$ in ${\cal G}_{\overline{X},\overline{Z_1}}$ and therefore $(Y(x,z_1) \ci z_1)_{{\cal G}(x,z_1)}$ by the definition of ${\cal G}(x,z_1)$.  Thus, if do-calculus Rule 3 precondition holds, po-calculus Rule 3 precondition holds.

We now prove the converse.
%The converse also holds.
If $(Y(x,z_1) \ci z_1)_{{\cal G}(x,z_1)}$ then $Z_1$ is not an ancestor of $Y$ in %${\cal G}$ (nor in
${\cal G}_{\overline{X},\overline{Z_1}}$.
Similarly if $(W(x,z_1) \ci z_1)_{{\cal G}(x,z_1)}$ then $Z_1$ is not an ancestor of $W$ in %${\cal G}$ (nor in
${\cal G}_{\overline{X},\overline{Z_1}}$.
Since $Z_1$ only has adjacent edges that are outgoing directed edges, this implies
$(Y,W \ci Z_1 \mid X)_{{\cal G}_{\overline{X},\overline{Z_1}}}$ holds.  Since semi-graphoid axioms hold for m-separation, this implies
$(Y \ci Z_1 \mid W,X)_{{\cal G}_{\overline{X},\overline{Z_1}}}$ holds.
%	$Y$ is also not an ancestor of $Z_1$ in ${\cal G}_{\overline{X},\overline{Z_1}}$ since ${\cal G}_{\overline{X},\overline{Z_1}}$ has no directed edges into $Z_1$. This implies that $Z_1$ and $Y$ are m-separated given $W,X$ in ${\cal G}_{\overline{X},\overline{Z_1}}$, since any m-connecting path given $W,X$ would have to be a path from $Z_1$ to $Y$ on which $W,X$ are colliders or descendants of colliders, contradicting the assumption that $W$ is a non-descendant of $Z_1$ and the definition of ${\cal G}_{\overline{X},\overline{Z_1}}$.
Finally,
$(Y(x,z_1) \ci Z_2(x,z_1) \mid W(x,z_1))_{{\cal G}(x,z_1)}$
holds if and only if
$(Y \ci Z_2 \mid W,X)_{{\cal G}_{\overline{X},\overline{Z_1}}}$
holds, by the definitions of ${\cal G}(x,z_1)$, ${\cal G}_{\overline{X},\overline{Z_1}}$, and m-separation. % in ${\cal G}(x,z_1)$.
\end{prf}

We now briefly demonstrate the soundness of the three rules of the po-calculus using only potential outcomes machinery and our background assumptions.
\begin{prop}
	Rules $1, 2,$ and $3$ are sound.
	\label{prop:sound}
\end{prop}
\begin{prf} Proposition \ref{prop:extended-cdag-markov} licenses deriving conditional independence statements corresponding to the graphical conditions in each rule. Then we have the following derivations:\\[2ex]
Rule $1$: $p(Y(x)|Z(x),W(x)) = p(Y(x)|W(x))$\\ by $Y(x) \ci Z(x) \mid W(x)$.\\ [2ex]
Rule $2$: $p(Y(x,z) | W(x,z)) = p(Y(x,z) | Z(x,z)=z, W(x,z)) = p(Y(x)|Z(x),W(x))$\\ by $Y(x,z) \ci Z(x,z) \mid W(x,z)$ and consistency.
%Rule $3$:
	\begin{align*}
	\text{Rule $3$: }&
	p(Y(x)|W(x)) = p(Y(x,z_1) | W(x,z_1))\\ & \hspace{4mm} \text{ since } Y(x,z_1), W(x,z_1) \ci z_1.
	%\color{red}\text{ and Cor. \ref{cor:simple-3}
%	Prop. \ref{prop:extended-cdag-markov}}
	%}}
	\\
	&\hspace{-1mm} = p(Y(x,z_1) | Z_2(x,z_1)=z_2, W(x,z_1))\\ & \hspace{4mm} \text{ since } Y(x,z_1) \ci Z_2(x,z_1) | W(x,z_1).
	%{\color{red}\text{ and Prop. \ref{prop:extended-cdag-markov}}}
	\\
	&\hspace{-1mm} = p(Y(x,z_1,z_2)|Z_2(x,z_1,z_2)=z_2 , W(x,z_1,z_2) )\\ & \hspace{4mm}
	\text{ by consistency.
	%{\color{red} by Prop. \ref{prop:consistency}}} 
	}\\
	&\hspace{-1mm} = p(Y(x,z) | Z_2(x,z)=z_2, W(x,z)) \\
	&\hspace{4mm} \text{ since $Y(x,z_1) \ci Z_2(x,z_1) \mid W(x,z_1)$,}\\
	& \hspace{4mm} \text{ $Z_2 \subseteq Z$, and so by Proposition \ref{prop:do-always-helps},}\\
	&\hspace{-1mm} = p(Y(x,z)|W(x,z))
	\end{align*} \end{prf}

The proof of Proposition \ref{prop:sound} has a number of interesting consequences.  First, the soundness of Rule $2$ follows by Rule $1$ and consistency.  Second, the soundness of Rule $3$ follows by applications of Rule $1$, Rule $2$, consistency, causal irrelevance, and intervention monotonicity.

Causal irrelevance, as used in the proof, is implied by m-separation statements in the SWIG ${\cal G}(x,z_1)$; however this property, like consistency, follows by (\ref{eqn:rec-sub}) alone and does not require 
any assumption regarding the distributions $p(V(a))$ for any $A \subseteq V$; specifically, (\ref{eqn:g}) is not required.
As a result the three rules of po-calculus, taken as a whole, are consequences of consistency and causal irrelevance, which hold in any recursive causal model, together with the SWIG Markov property for random variables in $V(a)$. (Intervention monotonicity follows from these.)
%As a result, the three rules of po-calculus taken as a whole are consequences of consistency and causal irrelevance, which hold in any recursive causal model, as well as the SWIG Markov property for random variables in $V(a)$ and its consequence, intervention monotonicity.
%, which hold in non-parametric structural equation models.
%^This makes it sound like causal models which are not NPSEM-IEs do not satisfy intervention monotonicity (though I realize that's not what is literally written).

%While working directly with consistency, causal irrelevance, and the global Markov property for % $p(V(a))$ and
%${\cal G}(a)$ is certainly possible for deductive problems involving counterfactuals, po-calculus suffices for many such problems in practice.  In fact, 
The proof of Proposition \ref{prop:sound} also implies that a simpler reformulation of po-calculus suffices without loss of generality. Specifically, this reformulation
replaces Rule $3$ by the following simpler rule (encoding causal irrelevance in graphical form):
%%A benefit of translating the do-calculus exactly into our potential outcomes formulation is that the do-calculus rules as stated have been shown to be sufficient for a wide class of possible derivations in causal inference. However, b
%By inspection of the previous proof we note that Rule $3$ can be replaced by the following, much simpler rule:
	\begin{align*}
	3^*\text{ : }& p(Y(x,z)) = p(Y(x)) \hspace{9mm} %\\
	%&
	% \hspace{20mm}
	\text{ if } (Y(x,z) \ci z)_{{\cal G}(x,z)}.
	%& \text{causal irrelevance,}
	\end{align*}
%which we may call ``fixed-node irrelevance,'' since it encodes an irrelevance relation like Prop. \ref{prop:causal-relevance} implied by separation in a SWIG. The soundness of Rule $3$ can be proved using Rule $3^*$ (the first equality in the above proof), and Rules $1$ and $2$ as well as consistency and Prop. \ref{prop:do-always-helps}, which are abstract properties of all recursive causal models. Thus, Rules $1$, $2$, and $3^*$ are in fact sufficient.

A benefit of translating the do-calculus exactly into our potential outcomes formulation is that the do-calculus rules as stated have been shown to be sufficient for a wide class of possible derivations on distributions expressible in terms of the $\doo$ operator
\cite{huang06do,shpitser06id}.
%(However, the do-calculus is not sufficient for deriving all implications of the assumed causal model; see the Supplement for discussion of this point.)
However, since we phrased the rules for arbitrary potential outcomes, they may be applied to causal contrasts not expressible in standard $\doo$ notation.  We illustrate this by applying these rules to mediation analysis.

\section{Path-Specific Effects and Extended Graphs}
The identification theory for path-specific effects generally proceeds by considering nested, path-specific potential outcomes. Fix a set of treatment variables $A$, and a subset of \emph{proper causal paths} $\pi$ from any element in $A$.  A proper causal path only intersects $A$ at the source node.  Next, pick a pair of value sets $a$ and $a'$ for elements in $A$.  For any $V_i \in V$, define the potential outcome $V_i(\pi,a,a')$ by setting $A$ to $a$ for the purposes of paths in $\pi$, and to $a'$ for the purposes of proper causal paths from $A$ to $Y$ not in $\pi$.  Formally, the definition is as follows, for any $V_i \in V$:
\begin{align}
\label{eqn:pse}
V_i(\pi, a, a') &\equiv a \text{ if }V_i \in A\\
V_i(\pi, a, a') &\equiv \notag \\
& \hspace{-8mm} V_i( \{ V_j(\pi, a, a') \mid V_j \in \Pa^{\pi}_i \}, \{ V_j(a') \mid V_j \in \Pa^{\overline{\pi}}_i \} )
\notag
\end{align}
where $V_j(a') \equiv a'$ if $V_j \in A$, and given by (\ref{eqn:rec-sub}) otherwise, 
$\Pa^{\pi}_i$ is the set of parents of $V_i$ along an edge which is a part of a path in $\pi$, and
$\Pa^{\overline{\pi}}_i$ is the set of all other parents of $V_i$.

A counterfactual $V_i(\pi, a, a')$ is said to be \emph{edge inconsistent} if counterfactuals of the form
$V_j(a_k, \ldots)$ and $V_j(a_k', \ldots)$ occur in $V_i(\pi, a, a')$, otherwise it is said to be \emph{edge consistent}.  It is well known that a joint distribution $p(V(\pi, a, a'))$ containing an edge-inconsistent counterfactual $V_i(\pi, a, a')$ is not identified in 
%${\cal C}_f({\cal G}(V))$, 
the functional causal model (nor weaker causal models)
with a corresponding graphical criterion on $\pi$ and ${\cal G}(V)$ called the `recanting witness' \cite{shpitser13cogsci,shpitser15hierarchy}.  For example, in Fig.~\ref{fig:triangle} (a), given $\pi = \{ C \to A \to Y \}$, $Y(\pi, c,c') \equiv Y(c', M(c', A(c')), A(c))$, while given $\pi = \{ A \to Y \}$,
$Y(\pi, a, a') \equiv Y(C, a, M(a',C))$.  Note that $Y(\pi, c, c')$ is edge inconsistent due to the presence of $A(c)$ and $A(c')$, while $Y(\pi, a, a')$ is edge consistent.

Counterfactuals defined by (\ref{eqn:pse}) form the basis for direct, indirect, and path-specific effects estimated in the mediation analysis literature.  There are generalizations where elements in $A$ are set to arbitrary values for different paths, under the name of \emph{path interventions} \cite{shpitser15hierarchy}.  Similarly, edge consistent counterfactuals $V(\pi, a, a')$ generalize to responses to \emph{edge interventions} \cite{shpitser15hierarchy}.  We do not discuss this further here in the interests of space, although the results presented below generalize without issue.  Note that edge consistent counterfactuals cannot, in general, be phrased in terms of the $\doo$ operator.

We have the following the result, proven in \cite{shpitser15hierarchy}.
\begin{thm}
	If $V(\pi, a, a')$ is edge consistent, then under the functional causal model for DAG $\mathcal{G}$,
	\begin{align}
	p(V(\pi, a, a')) = \prod_{i=1}^K p(V_i \mid a \cap \pa^{\pi}_i, a' \cap \pa^{\overline{\pi}}, \Pa^{\cal G}_i \setminus A).
	\label{eqn:edge-g}
	\end{align}
\end{thm}
As an example, the distribution $p(Y(\pi, a, a')) = p(Y(C,a,M(a',C)))$ of the edge consistent counterfactual in Fig.~\ref{fig:triangle} (a)
is identified as a marginal distribution derived from (\ref{eqn:edge-g}), specifically
$\sum_{C,M} p(Y \mid a,M,C) p(M \mid a',C) p(C)$.
The po-calculus as presented above may be applied to any sort of potential outcome, including nested potential outcomes representing path-specific effects. In the following, we exploit an equivalence between path-specific potential outcomes and standard potential outcomes defined from an \emph{extended graph} $\mathcal{G}^e$, which is constructed from $\mathcal{G}$ following \cite{robins10alternative}. This both simplifies complex nested potential outcome expressions and enables us to leverage a series of prior results to identify conditional PSEs.

%The idea is that complex, nested path-specific potential outcomes correspond to standard and simple potential outcomes on an appropriately defined extended variable set, and we can use the po-calculus deductive machinery to express these as functions of the observed data. 

Given an ADMG ${\cal G}(V)$, define for each $A_i \in A \subseteq V$ the set of variables
$A_i^{\Ch} \equiv \{ A_i^{j} \mid V_j \in \Ch_i \}$, and let $A^{\Ch} \equiv \bigcup_{A_i \in A} A_i^{\Ch}$.  We define the extended
graph of ${\cal G}(V)$, written ${\cal G}^e(V\cup A^{\Ch})$, as the graph with the vertex set $V \cup A^{\Ch}$, with edges of the form $A_i \to A_i^j \to V_j$ if and only if $A_i \to V_j$ is present in ${\cal G}(V)$, for $A_i \in A$, $V_j \in V$; furthermore, $V_i \leftrightarrow V_j$ in ${\cal G}^e(V\cup A^{\Ch})$ if and only if $V_i \leftrightarrow V_j$ is present for $V_i,V_j \in V$ in $\mathcal{G}(V)$.  As an example, the extended graph for the DAG in Fig.~\ref{fig:triangle} (a), with $A = V$, is shown in Fig.~\ref{fig:triangle} (c).  For conciseness, we will generally drop explicit references to vertices $V \cup A^{\Ch}$, and denote extended graph of ${\cal G}(V)$ by ${\cal G}^e$. Extended graphs as we define them here are straightforward generalizations of those presented in \cite{robins10alternative}, where they only consider ``node copies'' of a single ``treatment'' variable, whereas here extended graphs have ``copies'' corresponding to every parent-child relationship of a %selected
set of treatments $A$.
%\cite{robins10alternative} also limit their discussion to DAGs, whereas here we allow for ADMGs.

The edges $A_i \to A_i^j$ in $\mathcal{G}^e$ are understood to represent \emph{deterministic} relationships. More precisely, we associate a causal model with $\mathcal{G}^e$ as follows. For $\mathcal{G}$ we had associated a set of potential outcomes ${\mathbb V}$, and for $\mathcal{G}^e$ we have ${\mathbb V}^e$. For every $V_i(\pa_i) \in {\mathbb V}$, we let $V_i(\pa_i)$ be in ${\mathbb V}^e$.  Note that this is well-defined, since $V_i$ in ${\cal G}$ and ${\cal G}^e$ share the number of parents, and the parent sets for every $V_i$ share state spaces.  In addition, for every $A_i^j \in A^{\Ch}$, we let $A_i^j(a_i)$ for $a_i \in {\mathfrak X}_{A_i}$ be in ${\mathbb V}^e$. By assumption, every $A_i^j \in A^{\Ch}$ has a single parent $A_i$, and we further require that $p(A_i^j(a_i))$ is a deterministic density, with
$p(A_i^j(a_i) = a_i) = 1$. To fix intuitions, consider the example of Pearl's discussed in \cite{robins10alternative}. They consider an analysis where $A_i$ corresponds to smoking status, and affects hypertensive status $V_j$ as well as myocardial infarction status $V_{k}$ through nicotine $A_i^j$ and non-nicotine $A_i^k$ components respectively. The relationships $A_i \to A_i^j$ and $A_i \to A_i^k$ are deterministic relationships between smoking and exposure to nicotine/non-nicotine components. \cite{robins10alternative} go on to consider potential outcomes of the form $V_k(a_i^j,a_i^k)$ (where the ``node copies'' $A_i^j$ and $A_i^k$ are assigned to perhaps different values) inspired by a hypothetical intervention on the nicotine components of cigarette exposure that fixes non-nicotine components at some reference value (e.g., a new nicotine-free cigarette). In this case, the path-specific effect of smoking on outcome via nicotine components is easy to write down and identify, at the price of introducing new variables and deterministic relationships into the model.

We now show the following two results.  First, we show that an edge-consistent $V(\pi, a, a')$ may be represented without loss of generality by a counterfactual response to an intervention on a subset of $A^{\Ch}$ in ${\cal G}^e$ with the causal model defined above.  Second, we show that this response is identified by the same functional (\ref{eqn:edge-g}).

Given an edge consistent $V(\pi, a, a')$, define ${\cal G}^e$ via $A \subseteq V$.
%, and let
%$A^{\pi} \subseteq A^{\Ch}$ as $\{ A_i^j \text{ in }{\cal G}^e \mid A_i \in A \text{ in }{\cal G}(V) \}$. 
%$A^{\pi} \equiv A^{\Ch}$.
We define $a^{\pi}$ that assigns $a_i$ to $A_i^j \in A^{\Ch}$ if $A_i \to V_j$ in ${\cal G}(V)$ is in $\pi$, and assigns $a_i'$ to $A_i^j \in A^{\Ch}$ if $A_i \to V_j$ in ${\cal G}(V)$ is not in $\pi$.  The resulting set of counterfactuals $V(a^{\pi})$ is well defined in the model for
$\mathbb{V}^e$, %(V, A^{\Ch})$,
and we have the following result, proved in the Supplement.

\begin{prop}
Fix an element $p(\mathbb{V})$ in the causal model for a DAG ${\cal G}(V)$, and consider the corresponding element $p^e(\mathbb{V}^e)$ in the restricted causal model associated with a DAG ${\cal G}^e(V \cup A^{\Ch})$.  Then $p(V) = p^e(V \cup A^{\Ch})$ and $p(V(\pi, a, a')) = p^e(V(a^{\pi}))$.
%\begin{itemize}
%\item The observed data distributions in the two elements are the same up to determinism of $A$ and $A^{\Ch}$ in the model for ${\cal G}^e$.  That is, $p(V) = p^e(V \cup A^{\Ch})$, with $p^e(\{ A_i = A_i^j = a \mid A_i \in A \}) = 1$.
%\item $p(V(\pi, a, a')) = p^e(V(a^{\pi}))$.
%\end{itemize}
\label{prop:equiv}
\end{prop}

%\begin{prop}
%	$V(\pi, a, a')$ is equal to $V(a^{\pi})$ defined in the causal model for ${\cal G}^e$.
%\end{prop}
%\begin{prf}
%	Follows by definition of $a^{\pi}$, the fact that $p(A_i^j(a_i) = a_i) = 1$, and (\ref{eqn:rec-sub}).
%\end{prf}

\begin{cor}
	\label{cor:dag-ext}
	Given an extended DAG ${\cal G}^e$,
	\begin{align*}
	p(V(a^{\pi}))
	&= \prod_{i = 1}^K p^e(V_i \mid a^{\pi} \cap \pa_i, \Pa^{{\cal G}^e}_i \setminus A). %\\
	%&= \prod_{i = 1}^K p(V_i \mid a^{\pi} \cap \pa_i, \Pa^{\cal G}_i \setminus A).
	\end{align*}
\end{cor}
\begin{prf}
	This follows from Proposition \ref{prop:equiv}, and the fact that the functional in (\ref{eqn:edge-g}) in $p(V)$ is equal to
	$\prod_{i = 1}^K p^e(V_i \mid a^{\pi} \cap \pa_i, \Pa^{{\cal G}^e}_i \setminus A)$ in $p^e(V \cup A^{\Ch})$.
\end{prf}

In the causal models derived from DAGs with unobserved variables (e.g.,\ ${\cal G}(V \cup H)$), identification of distributions on potential outcomes
such as $p(V(a))$ or $p(V(\pi,a,a'))$ may be stated without loss of generality on the latent projection ADMG ${\cal G}(V)$.  %Note that not every such distribution is identified.  
A complete algorithm for identification of path-specific effects in hidden variable models was given in \cite{shpitser13cogsci} and presented in a more concise form in \cite{shpitser18e-id}.  We describe this form in detail in the Supplement.
We also note (and prove in the Supplement) that the latent projection and the extended graph operations commute.
%\begin{prop}
%Fix a DAG ${\cal G}(V \cup H)$, and let $A \subseteq V$.  Then ${\cal G}^e(V \cup A^{\Ch})$, the latent projection onto $V \cup A^{\Ch}$ of ${\cal G}^e(V \cup H \cup A^{\Ch})$ is equal to the extended graph ${\cal G}(V \cup A^{\Ch})^e$ applied to the latent projection ${\cal G}(V)$.
%\end{prop}

We now show that identification theory for $p(V(\pi,a,a'))$ in latent projection ADMGs ${\cal G}(V)$ may be restated, without loss of generality, in terms of identification of $p(V(a^{\pi}))$ in ${\cal G}^e(V \cup A^{\Ch})$.

%Fix a DAG ${\cal G}(V \cup H)$, and construct a DAG

\begin{prop}
For any $Y \subseteq V$,
$p(Y(\pi,a,a'))$ is identified in the ADMG ${\cal G}(V)$ if and only if $p(Y(a^{\pi}))$ is identified in the ADMG ${\cal G}^e(V, A^{\Ch})$.
Moreover if $p(Y(a^{\pi}))$ is identified, it is by the same functional as $p(Y(\pi,a,a'))$.
%, we have
%\[
%p(Y(a^{\pi})) = \sum_{Y^* \setminus Y} \prod_{D \in {\cal D}({\cal G}_{Y^*})} \left.
%\phi_{(V \cup A^{\Ch}) \setminus D}(p(V \cup A^{\Ch}); {\cal G}^e) \right|_{\tilde{a}_D},
%\]
%where $Y^* \equiv \an_{{\cal G}^e_{V \setminus A^{\Ch}}}(Y)$, and $\tilde{a}_D$ is defined to be the appropriate subset of $a^{\pi}$ associated with $\pa_{\cal G}(D) \cap A^{\Ch}$.
\label{prop:e-id-iff}
\end{prop}
Note that this Proposition is a generalization of Corollary \ref{cor:dag-ext} from DAGs to latent projection ADMGs.
The proof of this claim, and all claims in the next section, are given in the Supplement.

%\begin{prf}
%Argument structure.
%
%Identification:
%\begin{itemize}
%\item Show extended graph construction and latent projections commute.
%\item First assume there is no determinism in ${\cal G}^e$, and we get an identifying formula, but where kernels use ${\cal G}^e$ ``copy parents'' not ${\cal G}$ parents.
%\item Then note that under the recanting district criterion, each kernel piece in the resulting ID algorithm factorization only depends on a consistent set of treatments that do not have probability $0$ in observed data.  Further, that determinism allows us replace ${\cal G}^e$ parents with ${\cal G}$ parents.  (Note analogy with DAG proof).
%\end{itemize}
%Non-identification:
%\begin{itemize}
%\item Fix a witness for non-identifiability of $p(V(\pi,a,a'))$ in ${\cal G}(V)$.
%\item Translate the counterexample to ${\cal G}^e(V)$.
%\end{itemize}
%\end{prf}

%\begin{cor}
%	Every edge consistent $p(V(\pi, a, a'))$ in the causal model for $\mathcal{G}$ is equal to an interventional distribution $p(V(a^{\pi}))$ in the causal model for $\mathcal{G}^e$.  Moreover, both are identified via the same functional of $p(V)$.
%\end{cor}
%\begin{prf}
%
%\end{prf}
%
%{\color{red} [question: how do we know identified via the same functional when G is an ADMG, not DAG? do we need to argue that if identified by the g-functional in DAG, then still identified by the same functional after latent projection?]}

\section{Identification of Conditional PSEs}

Having established that we can identify path-specific effects by working with potential outcomes derived from the $\mathcal{G}^e$ model, we turn to the identification of conditional path-specific effects using the po-calculus. In \cite{shpitser06idc}, the authors present the conditional identification (IDC) algorithm for identifying quantities of the form $p(Y(x)|W(x))$ (in our notation), given an ADMG. Since conditional path-specific effects correspond to exactly such quantities defined on the extended model $\mathcal{G}^e$, we can leverage their scheme for our purposes. The idea is to reduce the conditional problem, identification of $p(Y(a^{\pi})|W(a^{\pi}))$, to an unconditional (joint) identification problem for which a complete identification algorithm already exists.

The algorithm has three steps: first, exhaustively apply Rule $2$ of the po-calculus to reduce the conditioning set as much as possible; second, identify the relevant joint distribution using Proposition \ref{prop:e-id-iff} and the complete algorithm in \cite{shpitser18e-id}; third, divide that joint by the marginal distribution of the remaining conditioning set to yield the conditional path-specific potential outcome distribution. The procedure is presented formally as Algorithm 1, with the subroutine corresponding to Proposition \ref{prop:e-id-iff} named PS-ID.
%, and rephrased in terms of interventions on an extended graph. This subroutine is described in detail in the Appendix.

%move this?
Note that we make use of SWIGs defined from extended graphs, e.g., ${\cal G}^e(a^{\pi},z)$. Beginning with ${\cal G}^e$ the SWIG ${\cal G}^e(a^{\pi},z)$ is constructed by the usual node-splitting operation: split nodes $Z$ and $A_i^j$ into random and fixed halves, where $A_i^j$ is has fixed copy $a$ if $A_i \to V_j$ in ${\cal G}(V)$ is in $\pi$, and $a'_i$ if $A_i \to V_j$ in ${\cal G}(V)$ is not in $\pi$. Relabeling of random nodes proceeds as previously described.

%\begin{algorithm}
%	\caption{PS-IDC($Y,a^{\pi},Z,\mathcal{G}$)
%		\label{alg:cpsid}}
%	\begin{algorithmic}[1]
%		\Statex \hspace{-6.5mm} \textbf{Output:} $p(Y(a^{\pi})|Z(a^{\pi}))$
%		%\Statex % for no line number
%		\State \textbf{if} $\exists Z' \in Z$ s.t.\
%		\Statex \hspace{8mm} $(Y(a^{\pi},z') \ci Z'(a^{\pi},z') \mid W(a^{\pi},z'))_{{\cal G}^e(a^{\pi},z')} $
%		\Statex \hspace{3mm} \textbf{return} PS-IDC($Y,a^{\pi} \cup z', Z \setminus Z', \mathcal{G}$)
%		\State \textbf{else} let $p'(Y(a^{\pi}),Z(a^{\pi})) \gets$ PS-ID($Y \cup Z, a^{\pi}, \mathcal{G}$)
%		\Statex \hspace{3mm} \textbf{return} $p'(Y(a^{\pi}),Z(a^{\pi})) / \sum_y p'(Y(a^{\pi}),Z(a^{\pi}))$
%	\end{algorithmic}
%\end{algorithm}

The following two results are adapted from \cite{shpitser06idc}; they are simply translated into potential outcomes and applied to extended graphs ${\cal G}^e$.
%We can adapt Shpitser and Pearl's (2008) proof of completeness as follows:

\begin{prop}
	If $(Y(x,z) \ci Z(x,z) \mid W(x,z))_{{\cal G}^e(x,z)}$ and $T\subseteq W$ then 
		$(Y(x,t) \ci T(x,t) \mid Z(x,t)$, $W_1(x,t))_{{\cal G}^e(x,t)}$
		if and only if 
		$(Y(x,z,t) \ci T(x,z,t) \mid W_1(x,z,t))_{{\cal G}^e(x,z,t)}$, where $W_1=W\setminus T$.
	\label{prop:util}
\end{prop}

%\begin{prop}
%	If $(Y(x,z) \ci Z(x,z) \mid W(x,z))_{{\cal G}^e(x,z)}$ then there are no m-connecting paths to $Y$ in neither $\mathcal{G}^e(x)$ given
%	$(Z(x),W(x))$ nor in $\mathcal{G}^e(x,z)$ given $W(x,z)$.
%	\label{prop:util}
%\end{prop}
%\begin{prf}
%	This is Shpitser and Pearl's (2008) Lemma 2 rephrased in terms of potential outcomes. The proof is the same.
%\end{prf}

\begin{cor} \label{cor:maximal}
	For any $\mathcal{G}^e(x)$ and any conditional distribution $p(Y(x)|W(x))$, there exists a unique maximal set $Z(x) = \{ Z_i(x) \in W(x) \mid p(Y(x)|W(x)) = p(Y(x,z_i)|W(x,z_i) \setminus \{ Z_i(x,z_i) \})\}$ such that Rule $2$ applies for $Z(x,z)$ in ${\cal G}^e(x,z)$ for $p(Y(x,z)|W(x,z))$.
\end{cor}
%\begin{prf}
%use Prop.~\ref{prop:util}.
%\end{prf}
%\begin{prf}
%	This is Shpitser and Pearl's (2008) Corollary 1 rephrased in terms of potential outcomes. The proof is the same.
%\end{prf}

%\begin{prop} \label{prop: separation}
%	For disjoint $Y,Z,W \subseteq V$, $(Y \ci Z \mid W)_{\mathcal{G}}$ if and only if $(Y \ci Z \mid W)_{\mathcal{G}^e}$.
%\end{prop}
%\begin{prf}
%	Follows from the construction of $\mathcal{G}^e$.
%\end{prf}
\begin{algorithm}
	\caption{PS-IDC($Y,a^{\pi},W,\mathcal{G}^e$)
		\label{alg:cpsid}}
	\begin{algorithmic}[1]
		\Statex \hspace{-6.5mm} \textbf{Input:} outcome $Y$, path-specific setting $a^{\pi}$,
		\Statex \hspace{4mm} conditioning set $W$, and graph ${\cal G}$
		\Statex \hspace{-6.5mm} \textbf{Output:} $p(Y(a^{\pi})|W(a^{\pi}))$
		%\Statex % for no line number
		\State \textbf{if} $\exists Z \in W$ s.t.\
		\Statex \hspace{8mm} $(Y(a^{\pi},z) \ci Z(a^{\pi},z) \mid W(a^{\pi},z))_{{\cal G}^e(a^{\pi},z)} $
		\Statex \hspace{3mm} \textbf{return} PS-IDC($Y,a^{\pi} \cup z, W \setminus Z, \mathcal{G}^e$)
		\State \textbf{else} let $p'(Y(a^{\pi}),W(a^{\pi})) \gets$ PS-ID($Y \cup W, a^{\pi}, \mathcal{G}^e$)
		\Statex \hspace{3mm} \textbf{return} $p'(Y(a^{\pi}),W(a^{\pi})) / \sum_y p'(Y(a^{\pi}),W(a^{\pi}))$
	\end{algorithmic}
\end{algorithm}
The following is similar to Theorem 6 in \cite{shpitser06idc}, but extended to path-specific queries in extended graphs. The proof is in the Supplement.
\begin{thm}
\label{thm:maximal}
	Let $p(Y(\pi,a,a') \mid W(\pi,a,a'))$ be a conditional path-specific distribution in the causal model for ${\cal G}$, and let
	$p(Y(a^{\pi}) \mid W(a^{\pi}))$ be the corresponding distribution in the extended causal model for ${\cal G}^e(V \cup A^{\Ch})$.
	Let $Z$ be the maximal subset of $W$ such that $p(Y(a^{\pi}) \mid W(a^{\pi})) = p(Y(a^{\pi},z) \mid W(a^{\pi},z) \setminus Z(a^{\pi},z))$.
	Then $p(Y(a^{\pi}) \mid W(a^{\pi}))$ is identifiable in ${\cal G}^e$ if and only if $p(Y(a^{\pi},z), W(a^{\pi},z) \setminus Z(a^{\pi},z))$ is identifiable in
	${\cal G}^e$.
\end{thm}
%\begin{prf}
%If $p(Y(x,z),W(x,z) \setminus Z(x,z))$ is identified in ${\cal G}^e$, our conclusion follows, since
%\begin{align*}
%p(Y(x)|W(x))
%&= p(Y(x,z)|W(x,z) \setminus Z(x,z))\\
%&= \frac{p(Y(x,z), W(x,z) \setminus Z(x,z))}{ p(W(x,z) \setminus Z(x,z)) }.
%\end{align*}
%If $p(Y(x,z),W(x,z) \setminus Z(x,z))$ is identified in ${\cal G}^e$, either $p(W(x,z))$ is identified or not.
%
%If $p(W(x,z))$ is identified $p(Y(x,z)|W(x,z) \setminus Z(x,z))$ is identified if and only if $p(Y(x,z),W(x,z) \setminus Z(x,z))$ is, and our conclusion follows.
%
%Assume $p(W(x,z))$ is not identified.
%\end{prf}
We then have by Corollary \ref{cor:maximal}, Theorem \ref{thm:maximal}, and completeness of the identification algorithm for path-specific effects \cite{shpitser18e-id}:
\begin{thm}
	Algorithm 1 is complete.
	\label{thm:complete}
\end{thm}
As an example, $p(Y(a,M(a')))$ is identified from $p(C,A,M,Y)$ in the causal model in Fig.~\ref{fig:example} (a), via
\[
%p(Y(a,M(a'))) =
\sum_{M} \frac{
\sum_{C} p(Y,M \mid a, C) p(C) \sum_{C} p(M \mid a', C) p(C)
}{
\sum_{C} p(M \mid a, C) p(C)
}.
\]
However $p(Y(a,M(a')) | C)$ is not identified, since PS-IDC concludes $p(Y(a,M(a')), C)$ must first be identified, and this joint distribution is not identified via results in \cite{shpitser13cogsci}. 
% and Theorem \ref{thm:complete}.
On the other hand, $p(Y(a,M(a')) | C)$ is identified from $p(C,A,M,Y)$ in a seemingly similar graph in Fig.~\ref{fig:example} (b), via
$\sum_{M} p(Y\mid M,a, C) p(M \mid a',C)$.

\section{Conclusion}

In this paper we introduced the potential outcomes calculus, a generalization of do-calculus that applies to arbitrary potential outcomes.  We have shown that potential outcome calculus is equivalent to Pearl's do-calculus for standard interventional quantities, and is a logical consequence of the properties of consistency and causal irrelevance, as well as the global Markov property associated with SWIGs. Finally, we used the potential outcomes calculus to give a sound and complete algorithm for conditional distributions defined on potential outcomes associated with path-specific effects. This algorithm may be viewed as a path-specific generalization of the identification algorithm for conditional interventional distributions in \cite{shpitser06idc}.

\section{Acknowledgments}
The authors would like to thank the American Institute of Mathematics for supporting this research via the SQuaRE program.
This project is sponsored in part by the 
National Institutes of Health grant R01 AI127271-01 A1, and the Office of Naval Research grants N00014-18-1-2760 and N00014-15-1-2672.
The authors would like to thank James M. Robins for helpful discussions.
%Blinded For Review

\small
\bibliographystyle{plain}
\bibliography{references}

\end{document}

% The file aaai.sty is the style file for AAAI Press 
% proceedings, working notes, and technical reports.
%
%\title{A Potential Outcomes Calculus for Identifying\\ Conditional Path-Specific Effects}
%\author{Authors\\
%Department of Computer Science\\
%Johns Hopkins University\\
%Baltimore, MD\\
%\texttt{\{emails\}@jhu.edu}
%}
%\maketitle

% If your paper is accepted and the title of your paper is very long,
% the style will print as headings an error message. Use the following
% command to supply a shorter title of your paper so that it can be
% used as headings.
%
\runningtitle{Supplement}

% If your paper is accepted and the number of authors is large, the
% style will print as headings an error message. Use the following
% command to supply a shorter version of the authors names so that
% they can be used as headings (for example, use only the surnames)
%
%\runningauthor{Surname 1, Surname 2, Surname 3, ...., Surname n}

\twocolumn[

\aistatstitle{Supplement For: A Potential Outcomes Calculus for Identifying\\ Conditional Path-Specific Effects}

\aistatsauthor{ Daniel Malinsky \And Ilya Shpitser \And Thomas Richardson }
%\aistatsauthor{ Author 1 \And Author 2 }

\aistatsaddress{ Johns Hopkins University \\ Department of Computer Science \\ Baltimore, MD USA\\ \texttt{malinsky@jhu.edu} \And Johns Hopkins University \\ Department of Computer Science \\ Baltimore, MD USA\\ \texttt{ilyas@cs.jhu.edu} \And University of Washington \\ Department of Statistics \\ Seattle, WA USA\\ \texttt{thomasr@uw.edu} } ]

Note that the Propositions with numbers $\leq 11$ are ones that appear in the body of the main paper, and Propositions with higher numbers are only stated here in the Supplement.

\section*{Utility Results}

First we define d-separation and m-separation for reference. 
Given two elements $V_i,V_j \in V$, and $X \subseteq V \setminus \{ V_i,V_j \}$, we say that a path
from $V_i$ to $V_j$ is blocked by $X$ if $*$$\to X_k \to$, $\gets X_k \gets$$*$, or $\gets X_k \to$ exists on the path, where $X_k \in X$, or if $*$$\to V_h \gets$$*$ exists on the path, where $V_h \not\in X$ and $\De_h \cap X = \emptyset$. $*$ stands for either an arrowhead or tail edge-mark, allowing for bidirected edges. We say $Y$ is m-separated from $Z$ given $X$ in ${\cal G}(V)$ if every path from an element of $Y$ to an element of $Z$ is blocked by $X$ in
${\cal G}(V)$. d-separation is the special case where all edges are directed.

\begin{proa}{4}%{\ref{prop:do-always-helps}}
	For any disjoint subsets $Y(a),Z(a),X(a)$ of $V(a)$ and a subset $a'$ of $a$,
	if $(Y(a), a' \ci Z(a) \mid X(a))_{{\cal G}(a)}$ then for any $A'' \supseteq A$,
	$(Y(a''), a' \ci Z(a'') \mid X(a''))_{{\cal G}(a'')}$.
\end{proa}

\begin{prf}
	Assume a m-connected path from an element in $Y$ or $a'$ to $Z(a)$ in ${\cal G}(a'')$.
	If this path does not intersect an element in $A'' \setminus A$, then it is also present in ${\cal G}(a)$.
	If this path does intersect $A'' \setminus A$, any element $A_i \in A''$ on this path cannot contain an outgoing edge on the path
	(since such edges do not exist in ${\cal G}(a'')$.  As a result, all edges on the path also exist in ${\cal G}(a)$.  Since the conditioning
	set is the same in both cases, the path is m-connected in ${\cal G}(a)$, which is a contradiction.
\end{prf}

\setcounter{prop}{11}
\begin{prop}
	Given a DAG ${\cal G}(V \cup H)$,
	${\cal G}(V)(a) = {\cal G}(a)(V)$.
	\label{eqn:swig-projection-commute}
\end{prop}
\begin{prf}
By definition, both graphs agree on the set of random and fixed vertices.
Note that ${\cal G}$ and ${\cal G}(a)$ have the same set of edges, and that $A \cap H = \emptyset$.
Consequently, any edge from $V_i$ to $V_j$ in ${\cal G}(V)(a)$ corresponds to a marginally d-connected path from $V_i$ to $V_j$ with all intermediate vertices in $H$ in ${\cal G}(V \cup H)$.  And similarly, such a path exists for any edge from $V_i$ to $V_j$ in ${\cal G}(a)(V)$.
This establishes the bijection between edges.
\end{prf}

Independence statements implied by d-separation on observed variable subsets of ${\cal G}(V \cup H)(a)$, for $A \subseteq V$ translate into m-separation statements of ${\cal G}(V)(a)$.

\begin{prop}
	For any disjoint subsets $Y(a),Z(a),X(a)$ of $V(a)$ and a subset $a'$ of $a$,
	\begin{align*}
	(Y(a), a' \ci Z(a) \mid X(a))_{{\cal G}(V \cup H)(a)} \Rightarrow \\
	(Y(a), a' \ci Z(a) \mid X(a))_{{\cal G}(V)(a)}.
	\end{align*}
	\label{lem:d-implies-m}
\end{prop}
\begin{prf}
This follows immediately from the fact that m-separation statements in a latent projection ADMG ${\cal G}(V)$ are in a one-to-one correspondence with d-separation statements in a DAG ${\cal G}(V \cup H)$ on $V$, and the SWIG global Markov property.
\end{prf}

\section*{A Complete Identification Algorithm For Path-Specific Counterfactual Distributions In Hidden Variable Causal Models}

Here we introduce a concise formulation of the complete identification algorithm for edge-consistent path-specific counterfactual distributions given in \cite{shpitser18e-id} via kernels, conditional graphs, and the fixing operation.

\subsection*{Kernels, Conditional Graphs, and Fixing} % the Nested Markov Model}
\label{subsec:kernels}

A kernel $q_V(V \mid W)$ is a mapping from ${\mathfrak X}_W$ to densities over $V$.  Given $A \subseteq V$, we define conditioning and marginalization in the usual way:
\begin{align*}
q_V(A | W) \equiv \sum_{V \setminus A} q_V(V | W); \hspace{0.1cm} q_V(V \setminus A | A, W) \equiv
\frac{
q_V(V | W)
}{
q_V(A | W)
}.
\end{align*}
A conditional graph ${\cal G}(V,W)$ is a graph with two types of vertices, random $V$ and fixed $W$, with the property that for any fixed vertex in $W$, its set of parents is empty.\footnote{Note that some elements of $V$ may have an empty parent set as well.}
We will consider conditional ADMGS (CADMGs), or conditional DAGs (CDAGs) as a special case.  A SWIG ${\cal G}(V(a))$ may be viewed as a conditional graph of the form ${\cal G}(V(a),a)$, where we denote the set of fixed vertices by $a$.

For a CADMG ${\cal G}(V,W)$, and $V_i \in V$, define
\begin{align*}
\Dis_i^{\cal G} &\equiv \{ V_j \mid V_j \leftrightarrow \ldots \leftrightarrow V_i \text{ in }{\cal G} \} & \text{ (district of $V_i$)}.
\end{align*}
Note that districts are only defined for, and may only contain, random vertices in $V$ not fixed vertices in $W$.
The set of districts in ${\cal G}$ is denoted by ${\cal D}({\cal G})$.

A vertex $V_i \in V$ in a CADMG ${\cal G}(V,W)$ is said to be fixable if $\De_i \cap \Dis_i = \emptyset$.  For such a vertex, define the operator $\phi_i({\cal G})$ that yields a new CADMG ${\cal G}(V \setminus \{ V_i \}, W \cup \{ V_i \})$, obtained by removing all edges with arrowheads into $V_i$, and keeping all other edges in ${\cal G}(V,W)$.

Given a CADMG ${\cal G}(V,W)$, and a kernel $q_V(V \mid W)$, if $V_i$ is fixable, define the operator $\phi_i(q_V; {\cal G})$ as yielding a
new kernel
\[
q_{V \setminus \{ V_i \}}(V \setminus \{ V_i \} | W \cup \{ V_i \}) \equiv
\frac{
q_V(V \mid W)
}{
q_V(V_i \mid \Mb_i^{\cal G}, W)
},
\]
where $\Mb_i^{\cal G}$, the \emph{Markov blanket} of $V_i$ in ${\cal G}$, is defined to be $\Dis_i^{\cal G} \cup \{ \Pa_j^{\cal G} \mid V_j \in \Dis_i^{\cal G} \}$.

A set of vertices $Z \subseteq V$ is said to be \emph{fixable} in ${\cal G}(V,W)$, if there exists a fixable sequence $Z_1, \ldots, Z_k$ on vertices in $Z$ such that $Z_1$ is fixable in ${\cal G}$, $Z_2$ is fixable in $\phi_1({\cal G})$, $Z_3$ is fixable in $\phi_2(\phi_1({\cal G}))$, and so on. Given a sequence $\alpha_Z$ for elements in $Z$, we define $\phi_{\alpha_Z}({\cal G})$ and $\phi_{\alpha_Z}(q_V;{\cal G})$ in the natural way by operator composition.  For any two valid fixing sequences $\alpha_Z,\beta_Z$ for a fixable set $Z$,
$\phi_{\alpha_Z}({\cal G}) = \phi_{\beta_Z}({\cal G})$.  Hence, for a fixable $Z$, we define $\phi_Z({\cal G})$ to mean ``fix elements in $Z$ in ${\cal G}$ by any fixable sequence.''

Given a CADMG ${\cal G}(V,W)$, if $Z \subseteq V$ is fixable, then $R \equiv V \setminus Z$ is called a reachable set.  A reachable set $R$ such that ${\cal D}(\phi_Z({\cal G}(V,W)))$ contains a single element is called \emph{intrinsic}.  If there exists a set of kernels
\[
\left\{ q_D(D | \Pa_D,W) \middle| D \text{ is intrinsic in }{\cal G}(V,W) \right\},
\]
where $\Pa_D \equiv \bigcup_{V_i \in D} \{ \Pa_i \setminus D \mid V_i \in D \}$, such that
for all fixable sets $Z$ in ${\cal G}(V,W)$, and all fixable sequences $\alpha_Z$, we have
\begin{align*}
\MoveEqLeft{\phi_{\alpha_Z}(q_V(V | W); {\cal G}(V,W)) =}\\
&\prod_{D \in {\cal D}(\phi_{Z}({\cal G}(V,W)))} q_D(D | \Pa_D, W),
\end{align*}
we say $q_V(V \mid W)$ is in the \emph{nested Markov model} of ${\cal G}(V,W)$.

For any such $q_V(V \mid W)$, it can be shown that for any fixable $Z$ in ${\cal G}(V,W)$, and any fixable sequences $\alpha,\beta$ for $Z$,
$\phi_{\alpha_Z}(q_V(V | W); {\cal G}(V,W)) = \phi_{\beta_Z}(q_V(V | W); {\cal G}(V,W))$.  As a result, we write
$\phi_{Z}(q_V(V | W); {\cal G}(V,W))$ to mean ``fix elements in $Z$ in $q_V(V | W)$ using any fixable sequence.''

Moreover, we have
\begin{align*}
\MoveEqLeft{\left\{ q_D(D | \Pa_D,W) \middle| D \text{ is intrinsic in }{\cal G}(V,W) \right\}}\\
\MoveEqLeft{= \{ \phi_{V \setminus D}(q_V(V\mid W); {\cal G}(V,W)) | \text{ is intrinsic in }{\cal G}(V,W) \}.}
\end{align*}

We have the following important results.
\begin{prop}
If $q_{V \cup H}(V \cup H | W)$ is in the Markov model for the CDAG ${\cal G}(V \cup H,W)$, then
$q_V(V | W) \equiv \sum_{H} q_{V\cup H}(V\cup H | W)$ is in the nested Markov model for the latent projection
CADMG ${\cal G}(V,W)$.
\label{prop:dag-implies-nested}
\end{prop}
\begin{prf}
This is shown in \cite{richardson17nested}.
\end{prf}

The complete algorithm for an edge-consistent $p(Y(\pi,a,a'))$ for $Y \subseteq V$ is stated as follows.
\begin{prop}
Let $Y^* \equiv \an_{{\cal G}_{V \setminus A}}(Y)$.  Then $p(Y(\pi,a,a'))$ is identified in ${\cal G}(V)$ if and only if
for every $D \in {\cal D}({\cal G}_{Y^*})$, $\pa_{\cal G}(D) \cap A$ are assigned to either a subset of $a$ or a subset of $a'$, and
$D$ is intrinsic in ${\cal G}(V)$.
Moreover, if $p(Y(\pi,a,a'))$ is identified, we have
\begin{align}
p(Y(\pi,a,a')) = \sum_{Y^* \setminus Y}  \prod_{D \in {\cal D}({\cal G}_{Y^*})} \left. \phi_{V \setminus D}(p(V); {\cal G}(V)) \right|_{\tilde{a}_D},
\label{eqn:id-e}
\end{align}
where $\tilde{a}_D$ is defined to be the appropriate subset of $a$ associated with $\pa_{\cal G}(D) \cap A$ if those elements are assigned by the definition of $Y(\pi,a,a')$, and the appropriate subset of $a'$ associated with $\pa_{\cal G}(D) \cap A$ otherwise.
\end{prop}
\begin{prf}
	This is shown in \cite{shpitser18e-id}.
\end{prf}

Note that the kernels $\phi_{V \setminus D}(.)$ are well defined by Proposition \ref{prop:dag-implies-nested}, since causal inference always starts with a causal model that implies a distribution that factorizes with respect to a (possibly hidden variable) DAG.

\subsection*{Remaining Proofs}

Now we turn to proving results related to Sections 5 and 6 in the main paper.

\begin{proa}{9}%{\ref{prop:equiv}}
Fix an element $p(\mathbb{V})$ in the causal model for a DAG ${\cal G}(V)$, and consider the corresponding element $p^e(\mathbb{V}^e)$ in the restricted causal model associated with a DAG ${\cal G}^e(V \cup A^{\Ch})$.  Then $p(V) = p^e(V,A^{\Ch})$ and $p(V(\pi, a, a')) = p^e(V(a^{\pi}))$. \label{prop:equiv}
\end{proa}

\begin{prf}
By definition of the causal model for ${\cal G}$, we have
\begin{align*}
p(V(\pi, a, a') = v) \!=\!\! \sum_{\epsilon_i : f_i(a_{\Pa^{\pi}_i}, a'_{\Pa^{\overline{\pi}}_i}, v_{\Pa_i \setminus A}) = v_i}
\!\!\! p(\epsilon_1, \ldots, \epsilon_k),
\end{align*}
where for each $V_i$, $\Pa^{\pi}_i$ is the subset of $\Pa_i \cap A$ with an edge from $\Pa_i$ to $V_i$ in $\pi$, and $\Pa^{\overline{\pi}}_i$
is the subset of $\Pa_i \cap A$ with an edge from $\Pa_i$ to $V_i$ not in $\pi$.
Similarly, by definition of the restricted causal model for ${\cal G}^e(V \cup A^{\Ch})$, we have
\begin{align*}
p^e(V(a^{\pi}) = v) = \sum_{\epsilon_i : f_i(a^{\pi}_{\Pa_i \cap A^{\Ch}}, v_{\Pa_i \setminus A}) = v_i}
p(\epsilon_1, \ldots, \epsilon_k).
\end{align*}
The equivalence follows immediately.  Note that the same argument establishes $p(V) = p^e(V, A^{\Ch})$, by letting $\pi$ be the empty set of paths, and $A = \emptyset$.
\end{prf}

\begin{prop}
Assume there exists elements $p_1({\mathbb V}),p_2({\mathbb V})$ in the causal model for ${\cal G}$ such that
$p_1(V) = p_2(V)$, but $p_1(V(\pi,a,a')) \neq p_2(V(\pi,a,a'))$.  Then $p(V(a^{\pi}))$ is not identified in the restricted causal model for
${\cal G}^e(V \cup A^{\Ch})$.
\label{cor:non-id}
\end{prop}
\begin{prf}
Follows immediately by Proposition 9.%\ref{prop:equiv}.
\end{prf}

We state formally our claim in the main paper that the latent projection and extended graph operations commute.

\begin{prop}
	Fix a DAG ${\cal G}(V \cup H)$, and let $A \subseteq V$.  Then ${\cal G}^e(V \cup A^{\Ch})$, the latent projection onto $V \cup A^{\Ch}$ of ${\cal G}^e(V \cup H \cup A^{\Ch})$ is equal to the extended graph ${\cal G}(V \cup A^{\Ch})^e$ applied to the latent projection ${\cal G}(V)$.
\end{prop}
\begin{prf}
	%\color{red}
	By definition, the two graphs have the same vertices. That the two graphs share the same edges follows from the definition of ${\cal G}^e$, which stipulates that the only edge into each variable in $A^{\Ch}$ is from the corresponding variable in $A$, i.e., there are no directed paths from any $H$ into any element of $A^{\Ch}$ not through some element of $A$. So, all bidirected edges induced by the latent projection operation are between vertices in $V$, which are shared between the two graphs.
\end{prf}

%Finally, we turn to proving the remaining claims stated in the main paper.

\begin{proa}{10}%{\ref{prop:e-id-iff}}
For any $Y \subseteq V$,
$p(Y(\pi,a,a'))$ is identified in the ADMG ${\cal G}(V)$ if and only if $p(Y(a^{\pi}))$ is identified in the ADMG ${\cal G}^e(V, A^{\Ch})$.
Moreover if $p(Y(a^{\pi}))$ is identified, we have
\begin{align}
\notag
\MoveEqLeft{p^e(Y(a^{\pi})) =}\\
 &\sum_{Y^* \setminus Y} \prod_{D \in {\cal D}({\cal G}^e_{Y^*})} \left.
\phi_{(V \cup A^{\Ch}) \setminus D}(p^e(V, A^{\Ch}); {\cal G}^e) \right|_{\tilde{a}_D}
\label{eqn:id-e-2}
\end{align}
where $Y^* \equiv \an_{{\cal G}^e_{V \setminus A^{\Ch}}}(Y)$, and $\tilde{a}_D$ is defined to be the appropriate subset of $a^{\pi}$ associated with $\pa_{\cal G}(D) \cap A^{\Ch}$.
\end{proa}

\begin{prf}
Assume $p(Y(\pi,a,a'))$ is identified in ${\cal G}(V)$ via (\ref{eqn:id-e}).  The conclusion follows from Proposition \ref{prop:equiv}, and the 		fact that the functional in (\ref{eqn:id-e}) in $p(V)$ is equal to
	(\ref{eqn:id-e-2}) in $p^e(V, A^{\Ch})$.

Assume $p(Y(\pi,a,a'))$ is not identified, and fix a witness of this fact, which is either a hedge or a district with a recanting set of parents in $A$.  If the witness is a hedge, the construction in \cite{shpitser06id} yields $p_1({\mathbb V})$ and $p_2({\mathbb V})$, such that
$p_1(V) = p_2(V)$, but $p_1(Y(\pi, a, a')) \neq p_2(Y(\pi, a, a'))$
If the witness is a recanting district, the construction in \cite{shpitser13cogsci}, described also in \cite{shpitser18e-id}, yields
$p_1({\mathbb V})$ and $p_2({\mathbb V})$, such that $p_1(V) = p_2(V)$, but $p_1(Y(\pi, a, a')) \neq p_2(Y(\pi, a, a'))$.
In both cases, this immediately implies the conclusion by Corollary \ref{cor:non-id}.
\end{prf}

\setcounter{prop}{10}
\begin{prop}
	If $(Y(x,z) \ci Z(x,z) \mid W(x,z))_{{\cal G}^e(x,z)}$ and $T\subseteq W$ then 
	$(Y(x,t) \ci T(x,t) \mid Z(x,t)$, $W_1(x,t))_{{\cal G}^e(x,t)}$
	if and only if 
	$(Y(x,z,t) \ci T(x,z,t) \mid W_1(x,z,t))_{{\cal G}^e(x,z,t)}$, where $W_1=W\setminus T$.
	\label{prop:util}
\end{prop}
\begin{prf}
	The set of possible d-connecting paths from $Y(x,z,t)$ to $T(x,z,t)$ in ${\cal G}^e(x,z,t)$ is a subset of the set of possible d-connecting paths from $Y(x,t)$ to $T(x,t)$ in ${\cal G}^e(x,t)$. For any such path that exists in both graphs, if it is blocked by $W_1(x,t)$ in ${\cal G}^e(x,t)$, it will be blocked by $W_1(x,z,t)$ in ${\cal G}^e(x,z,t)$.  If it is blocked by $Z(x,t)$ in ${\cal G}^e(x,t)$, the path will be blocked in ${\cal G}^e(x,z,t)$ by construction of ${\cal G}^e(x,z,t)$.  If it is blocked by collider without $Z(x,t),W_1(x,t)$ descendants in ${\cal G}^e(x,t)$, the same will remain true in ${\cal G}^e(x,z,t)$.  Thus, if $(Y(x,t) \ci T(x,t) \mid Z(x,t)$, $W_1(x,t))_{{\cal G}^e(x,t)}$, then $(Y(x,z,t) \ci T(x,z,t) \mid W_1(x,z,t))_{{\cal G}^e(x,z,t)}$.
	
	Now, assume for contradiction, $(Y(x,z,t) \ci T(x,z,t) \mid W_1(x,z,t))_{{\cal G}^e(x,z,t)}$, but
	$(Y(x,t) \not\ci T(x,t) \mid Z(x,t)$, $W_1(x,t))_{{\cal G}^e(x,t)}$, with a witnessing d-connecting path from some $Y_1(x,t)$ to some $T_1(x,t)$.  If this path is not a possible d-connecting path in ${\cal G}^e(x,z,t)$, it must contain a non-collider through an element of $Z$, and thus is blocked by $Z(x,t)$ in ${\cal G}^e(x,t)$.  If this path is a possible d-connecting path in ${\cal G}^e(x,z,t)$ it must be blocked by a collider which contains no descendants in $W_1(x,z,t)$ in ${\cal G}^e(x,z,t)$, but remains open due to this collider containing descendants in $Z(x,t)$ in ${\cal G}^e(x,t)$.
	
	But this implies the existence of a d-connecting path in ${\cal G}^e(x,t)$ from an element $Y_1(x,t)$ in
	$Y(x,t)$ to an element $Z_1(x,t)$ in $Z(x,t)$ given $W_1(x,t)$, and thus also given $W(x,t)$ (since no element in $T(x,t)$ will block this path by construction).  Since we can choose $Z_1(x,t)$ to be the closest element in $Z(x,t)$ to $Y_1(x,t)$ involved in the witnessing path, we obtain that $Y_1(x,z) \not\ci Z_1(x,z) \mid W(x,z)$, which is a contradiction.
\end{prf}

\begin{cora}{2}%{\ref{cor:maximal}}
	For any $\mathcal{G}^e(x)$ and any conditional distribution $p(Y(x)|W(x))$, there exists a unique maximal set $Z(x) = \{ Z_i(x) \in W(x) \mid p(Y(x)|W(x)) = p(Y(x,z_i)|W(x,z_i) \setminus \{ Z_i(x,z_i) \})\}$ such that Rule $2$ applies for $Z(x,z)$ in ${\cal G}^e(x,z)$ for $p(Y(x,z)|W(x,z))$.
\end{cora}

\begin{prf}
	Fix two maximal sets $Z_1(x)$ and $Z_2(x)$ such that Rule $2$ applies for $Z(x,z)$ in ${\cal G}^e(x,z)$ for $p(Y(x,z)|W(x,z))$. If $Z_1(x) \neq Z_2(x)$, fix $T(x) \in Z_1(x) \setminus Z_2(x)$. By the previous proposition, Rule $2$ applies for $Z_2(x) \cup T(x)$, contradicting our assumption.
\end{prf}
\begin{thma}{2}%{\ref{thm:maximal}}
	Let $p(Y(\pi,a,a') \mid W(\pi,a,a'))$ be a conditional path-specific distribution in the causal model for ${\cal G}$, and let
	$p(Y(a^{\pi}) \mid W(a^{\pi}))$ be the corresponding distribution in the extended causal model for ${\cal G}^e(V \cup A^{\Ch})$.
	Let $Z$ be the maximal subset of $W$ such that $p(Y(a^{\pi}) \mid W(a^{\pi})) = p(Y(a^{\pi},z) \mid W(a^{\pi},z) \setminus Z(a^{\pi},z))$.
	Then $p(Y(a^{\pi}) \mid W(a^{\pi}))$ is identifiable in ${\cal G}^e$ if and only if $p(Y(a^{\pi},z), W(a^{\pi},z) \setminus Z(a^{\pi},z))$ is identifiable in ${\cal G}^e$.
\end{thma}

\begin{prf}
The proof strategy follows that of the completeness argument in \cite{shpitser06idc}.  We expand the argument here to be more transparent.  In addition, we must handle an additional case of non-identifiability that arises in mediation problems, that has to do with structures called \emph{recanting districts} in \cite{shpitser13cogsci}.

If $p(Y(a^{\pi},z),W(a^{\pi},z) \setminus Z(a^{\pi},z))$ is identified in ${\cal G}^e$, then $p(Y(a^{\pi}) \mid W(a^{\pi}))$ is identifiable in ${\cal G}^e$ since
\begin{align*}
p(Y(a^{\pi})|W(a^{\pi}))
&= p(Y(a^{\pi},z)|W(a^{\pi},z) \setminus Z(a^{\pi},z))\\
&= \frac{p(Y(a^{\pi},z), W(a^{\pi},z) \setminus Z(a^{\pi},z))}{ p(W(a^{\pi},z) \setminus Z(a^{\pi},z)) }.
\end{align*}
Now assume $p(Y(a^{\pi},z),W(a^{\pi},z) \setminus Z(a^{\pi},z))$ is not identified in ${\cal G}^e$. Either $p(W(a^{\pi},z))$ is identified or not.
If $p(W(a^{\pi},z))$ is identified, $p(Y(a^{\pi},z)|W(a^{\pi},z) \setminus Z(a^{\pi},z))$ is identified if and only if $p(Y(a^{\pi},z),W(a^{\pi},z) \setminus Z(a^{\pi},z))$ is. Since the latter is false by assumption, our conclusion follows.

Assume $p(W(a^{\pi},z))$ is not identified.  Let $\tilde{a} = a \cup z$, and $\tilde{\pi}$
be the set comprised of $\pi$ and all outgoing directed edges from elements in $Z$.  Then the distribution
$p(W(a^{\pi},z))$ is equal to $p(W(\tilde{a}^{\tilde{\pi}}))$, which in turn is equivalent to $p(W(\tilde{\pi},\tilde{a},a'))$.

$p(W(\tilde{\pi},\tilde{a},a'))$ could fail to be identified in the causal model for ${\cal G}$ for two reasons.  Either there could exist a hedge structure \cite{shpitser06id} for $p(W(a))$, or there could exist a recanting district structure \cite{shpitser13cogsci} in
${\cal D}({\cal G}_{W^*})$, where $W^* \equiv \An_W^{{\cal G}_{V \setminus A}}$.  We consider these cases in turn.

If there exists a hedge structure, %then $p(W(\tilde{\pi},\tilde{a},a'))$ because $p(W(a))$ is not identified.
fix a district $D$ in ${\cal D}({\cal G}_{W^*})$, where $W^* \equiv \An^{{\cal G}_{V \setminus A}}_{W}$, such that
there is a larger district $D'$ containing $D$ that forms the hedge structure with $D$.  Further, find the minimal subset $W'$ of $W$ such that the set of all childless vertices in the hedge structure (contained in $D'$) is in $\An^{{\cal G}_{V \setminus A}}_{W'}$.  Let $H$ be the smallest set of vertices that contains $W'$, $D'$, and such that the set of childless vertices in the hedge structure is in $\An^{{\cal G}_{H}}_{W'}$.

Assume without loss of generality that each vertex in ${\cal G}_H$ has at most one child.  We construct elements $p_1({\mathbb H})$ and $p_2({\mathbb H})$ in the causal model in ${\cal G}_H$ as follows.  In $p_1({\mathbb H})$ each structural equation is a bit parity function of the parents, and each bidirected arc corresponds to a binary latent common parent where each such latent is involved in precisely two functions.  Moreover, each such latent variable $\epsilon_{ij}$ that is a parent of $V_i$ and $V_j$ is drawn from a uniform distribution $p(\epsilon_{ij})$.
In $p_2({\mathbb H})$ the same is true, except no element in $D' \setminus D$ is involved in the structural equation for any element in $D$, and no $\epsilon_{ij}$ that is a parent of an element in $D' \setminus D$ and an element in $D$ exists.  It has been shown in \cite{shpitser06id} that if $p_1({\mathbb H})$ and $p_2({\mathbb H})$ are constructed in this way, they induce $p_1(H),p_1(W'({a}_{H \cap A}))$ and $p_2(H),p_2(W'({a}_{H \cap A}))$ respectively, such that $p_1(H) = p_2(H)$ (i.e., the induced observational distributions are the same), but $p_1(W'({a}_{H \cap A})) \neq p_2(W'({a}_{H \cap A}))$ (i.e., the induced potential outcome distributions are distinct).

Specifically, let $R$ be the set of childless vertices in ${\cal G}_{D'}$.  Then it has been shown that $p_1(D') = p_2(D')$ is a distribution uniform over any assignment to $D'$ such that the number of $1$ values in $R$ is even.  At the same time, $p_1(R(a_{H \cap A}))$ is a uniform distribution over assignments with even number of $1$ values, while $p_2(R(a_{H \cap A}))$ is a uniform distribution.
Since each element in $H \setminus R$ has a single parent in ${\cal G}_H$, the bit parity function for those elements simply reduces to the identity function.  Note that more general structural equations suffice for the argument, as long as the linear transformation that maps $p(D'(a_{H \cap A}))$ to $p(W'(a_{H \cap A})) \equiv \sum_{D'} p(W'(a_{H \cap A}) \mid D'(a_{H \cap A})) p(D'(a_{H \cap A}))$ is one to one.

%We extend ${\cal G}_H$ with vertices on
Consider a path $\pi$ in ${\cal G}(a)$ from some element $W_i$ in $W'$ to an element $Y_j$ in $Y$, such that $W_i$ is m-connected to $Y$ given $W$, and the edge on the path adjacent to $W_i$ has an arrowhead into $W_i$ (Pearl called such paths \emph{backdoor paths}).  Such a path must exist by construction of $W$.  In addition, consider the smallest subset $W''$ of $W$ such that $W_i$ is m-connected to $Y_j$ given $W''$ in ${\cal G}(a)$.  Pick the smallest set $H'$ containing $H$ such that the above m-connection statement holds in
${\cal G}(a)_{H'}$. We now extend $p_1({\mathbb H})$ and $p_2({\mathbb H})$ to $p_1({\mathbb H}')$ and $p_2({\mathbb H}')$ to show
$p(Y_j(a_{H' \cap A}) \mid W'(a_{H' \cap A}))$ is not identified.

%{\color{blue}For the moment, we assume $H \subset H'$.}
We have three base cases.  The first case assumes the first node $Z_j$ on $\pi$ not in $H$ is a parent of an element $Z_i$ in $H$.
Let the structural equation corresponding to $Z_i$ be the bit parity function of all its parents in ${\cal G}_{H'}$, including $Z_j$ in both $p_1({\mathbb H}')$ and $p_2({\mathbb H}')$, and let $p(Z_j)$ be the uniform distribution on a binary variable.

In this case, the observed data distributions are $p_1(H \mid Z_j) p_1(Z_j)$ and $p_2(H \mid Z_j) p_2(Z_j)$.  $p_1(Z_j) = p_2(Z_j)$ by construction.  Next, note that $p_1(H \mid Z_j=0) = p_2(H \mid Z_j=0)$ equal to the distributions $p_1(H) = p_2(H)$ given in the previous construction.  Specifically these distributions are uniform on all assignments to $R$ with an even number of $1$ values.  By symmetry,
$p_1(H \mid Z_j=1) = p_2(H \mid Z_j=1)$, with the distributions being uniform on all assignments to $R$ with an odd number of $1$ values.
By above construction and results in \cite{shpitser06id}, $p_1(H(a_{H \cap A}) \mid Z_j(a_{H \cap A}) = 0) = p_1(H\mid Z_j= 0)$,
while $p_2(H(a_{H \cap A}) \mid Z_j(a_{H \cap A}) = 0)$ is a uniform distribution.  Since $p_1(Z_j(a_{H \cap A})) = p_1(Z_j) = p_2(Z_j) = p_2(Z_j(a_{H \cap A}))$, we have that
$p_1(Z_j(a_{H \cap A}), R(a_{H\cap A}))$ only has positive probability if the number of $1$ values in $\{ Z_j \} \cup R$ is even, while
$p_2(Z_j(a_{H \cap A}), R(a_{H\cap A}))$ is a uniform distribution.  This implies
$p_1(Z_j(a_{H \cap A}) = 0 \mid R(a_{H\cap A}) = 0) = 1$, while
$p_2(Z_j(a_{H \cap A}) = 0 \mid R(a_{H\cap A}) = 0) < 1$, which establishes the base case.

The second case assumes the first node $Z_j$ on $\pi$ not in $H$ is a child of an element $Z_i$ in $H$.  We also consider the third case  where $Y_j \in H$, here by letting $Y_j = Z_i$.  If $p(Z_j(a_{H' \cap A}) \mid W'(a_{H' \cap A}))$ (or $p(Y_j(a_{H' \cap A}) \mid W'(a_{H' \cap A}))$) is not identified, we are done.  Otherwise, we assume $p(Z_j(a_{H' \cap A}) \mid W'(a_{H' \cap A}))$ is identified.  Consider the edge subgraph ${\cal G}'_{H'}$ of ${\cal G}_{H'}$ that lacks the outgoing directed edges from $Z_i$ within $H$.

Since the childless vertices in the hedge structure are in $\An^{{\cal G}_{H}}_{W'}$, if $Z_i$ is not in the hedge structure in $H$, it must be on a directed path in ${\cal G}_{H}$ from some childless vertex in the hedge structure to an element of $W'$.  Since we assumed each vertex in ${\cal G}_H$ has at most one child, removing the outgoing arrow from $Z_i$ in ${\cal G}_{H'}$ results in 
{${\cal G}'_{H'}$ containing the hedge structure for $p(Z_j(a_{H' \cap A}), W''(a_{H' \cap A}))$, where $W'' = W' \setminus \{ W_i \}$ and $W_i$ is $W' \cap \De_{{\cal G}_{H'}}(Z_j)$}.

If $p(W''(a_{H' \cap A}))$ is identified, we are done, since we established the base case where $p(Z_j(a_{H' \cap A}) \mid W''(a_{H' \cap A}))$ is not identified.  If $p(W''(a_{H' \cap A}))$ is not identified, note that $W''$ is a strictly smaller set then $W'$, and we restart the base case argument, finding a hedge or a recanting district for this smaller set, constructing a new set $H$, and a new backdoor path to an element in $Y$.  Since the new subset of $W$ is strictly smaller, we can only do this a finite number of times before encountering another base case.
%{\color{red}What about $Y \in H$?}

If $Z_i$ is in the hedge structure in $H$, then the resulting graph ${\cal G}'_{H'}$ contains a hedge structure for $p(Z_j(a_{H' \cap A}), W'(a_{H' \cap A}))$ with the set of childless vertices of the previous hedge and also $Z_i$ (since it is now childless in $H$).
Given the hedge construction above, we have
{$p_1(Z_j(a_{H' \cap A}) = 0 \mid W'(a_{H' \cap A}) = 0) < 1$, while
$p_2(Z_j(a_{H' \cap A}) = 0 \mid W'(a_{H' \cap A}) = 0) = 1$}, and we are done.

We now consider the inductive cases on the path $\pi$.  Consider $Z_{k}$ and $Z_{k+1}$ on the path, where $Z_{k+1}$ is closer to 
$Y_j$ on the path.  We have the following cases.

If $Z_{k+1}$ is a parent of $Z_{k}$, or $Z_{k+1}$ is a child of $Z_{k}$, then in ${\cal G}'_{H'}$:
\begin{small}
\begin{align*}
&p_1(Z_{k+1}(a_{H' \cap A}) | W'(a_{H' \cap A}))
= \hspace{5cm} \\
 &\sum_{Z_{k}} p_1(Z_{k}(a_{H' \cap A}) | W'(a_{H' \cap A})) p_1(Z_{k+1}(a_{H' \cap A}) | Z_{k}(a_{H' \cap A}))\\
&p_2(Z_{k+1}(a_{H' \cap A}) | W'(a_{H' \cap A}))
= \hspace{5cm}\\
 &\sum_{Z_{k}} p_2(Z_{k}(a_{H' \cap A}) | W'(a_{H' \cap A})) p_2(Z_{k+1}(a_{H' \cap A}) | Z_{k}(a_{H' \cap A})).
\end{align*}
\end{small}
%{\color{red}[is the above right? think you had a typo before]} %{\color{blue}Was wrong before, thanks for fix, what you have is right.}
Let
\begin{small}
\[
p_1(Z_{k+1}(a_{H' \cap A}) | Z_{k}(a_{H' \cap A})) = p_2(Z_{k+1}(a_{H' \cap A}) | Z_{k}(a_{H' \cap A})).
\]
\end{small}
Then we have
\begin{small}
\[
p_1(Z_{k+1}(a_{H' \cap A}) | W'(a_{H' \cap A})) \neq p_2(Z_{k+1}(a_{H' \cap A}) | W'(a_{H' \cap A}))
\]
\end{small}
if and only if
\begin{small}
\[
p_1(Z_{k}(a_{H' \cap A}) | W'(a_{H' \cap A})) \neq p_2(Z_{k}(a_{H' \cap A}) | W'(a_{H' \cap A})).
\]
\end{small}
These latter distributions are not equal in $p_1(\mathbb{H}')$ and $p_2(\mathbb{H}')$ by the inductive hypothesis.
% because of the hedge structure for $p(Z_j(a_{H' \cap A}) , W''(a_{H' \cap A}))$}.

%This may be viewed as a linear map, which may trivially be made one to one by choosing 
%$p_1(Z_{k+1}(a_{H' \cap A}) | Z_{k}(a_{H' \cap A})) = p_2(Z_{k+1}(a_{H' \cap A}) | Z_{k}(a_{H' \cap A}))$ to be a full column rank matrix
%with appropriate normalization properties (recall that for binary $Z_k,Z_{k+1}$, this distribution is a 2 by 2 matrix).

If $Z_{k+1}$ is a sibling of $Z_{k}$, we repeat the above two cases, since this case may be rephrased without loss of generality in terms of an unobserved variable $H_k$ that is a parent of both $Z_{k+1}$ and $Z_{k}$.

If $Z_{k+1}$ and $Z_{k}$ are both parents of a variable $C_{k}$ which is an ancestor of element $W_k$ in $W'$,
we have:
\begin{align*}
\MoveEqLeft{p_1(Z_{k+1}(a_{H' \cap A}) | W_k(a_{H' \cap A}), W'(a_{H' \cap A})) =}\\
\MoveEqLeft{\kern10pt\sum_{Z_{k}} p_1(Z_{k+1}%(a_{H' \cap A})
| W_k, Z_{k}%(a_{H' \cap A})
) \times} \\
\MoveEqLeft{\frac{p_1(W_k | Z_k) }{\sum_{Z_k} p_1(W_k | Z_k) p_1(Z_k | W'(a_{H' \cap A}))} p_1(Z_k | W'(a_{H' \cap A}))}\\[8pt]
\MoveEqLeft{p_2(Z_{k+1}(a_{H' \cap A}) | W_k(a_{H' \cap A}), W'(a_{H' \cap A})) =}\\
\MoveEqLeft{\kern10pt\sum_{Z_{k}} p_2(Z_{k+1}%(a_{H' \cap A}) 
| W_k, Z_{k}%(a_{H' \cap A})
) \times }\\
\MoveEqLeft{\frac{p_2(W_k | Z_k) }{\sum_{Z_k} p_2(W_k | Z_k) p_2(Z_k | W'(a_{H' \cap A}))} p_2(Z_k | W'(a_{H' \cap A}))}
\end{align*}
Assume $W_k = C_k$.
We must choose
\[
p_1(Z_{k+1},W_k \mid Z_k) = p_2(Z_{k+1},W_k \mid Z_k)
\]
such that
\begin{align*}
p_1(Z_{k+1}(a_{H' \cap A}) | W_k(a_{H' \cap A}), W'(a_{H' \cap A})) \neq \\
p_2(Z_{k+1}(a_{H' \cap A}) | W_k(a_{H' \cap A}), W'(a_{H' \cap A}))
\end{align*}
if
\[
p_1(Z_k | W'(a_{H' \cap A})) \neq p_2(Z_k | W'(a_{H' \cap A})),
\]
(which is true by the inductive hypothesis).

For a fixed $W_k$, we have 5 degrees of freedom: $p(Z_{k+1})$,
$p(W_k | Z_{k+1}, Z_k)$,
$p(W_k | Z_{k+1}, 1-Z_k)$,
$p(W_k | 1-Z_{k+1}, Z_k)$, and
$p(W_k | 1-Z_{k+1}, 1-Z_k)$.

%\color{red}
It suffices to specify these in such a way that the linear mapping induced by
\[
p(Z_{k+1} \mid W_k, Z_k) = \frac{ p(Z_{k+1}) p(W_k \mid Z_{k+1}, Z_k) }{ \sum_{Z_{k+1}}  p(Z_{k+1}) p(W_k \mid Z_{k+1}, Z_k) }
\]
is a one-to-one mapping, %full column rank matrix,
and for some $W_k$,
$c = p_1(W_k \mid Z_k) = p_2(W_k \mid Z_k)$, and
$k = p_1(W_k \mid 1 - Z_k) = p_2(W_k \mid 1 - Z_k)$ are chosen such that
\[
\frac{p_1(Z_k | W'(a_{H' \cap A}))}{p_2(Z_k | W'(a_{H' \cap A}))} \neq 
\frac{k + p_1(Z_k | W'(a_{H' \cap A}))(c - k)}{k + p_2(Z_k | W'(a_{H' \cap A}))(c - k)},
\]
for some $p_1(Z_k | W'(a_{H' \cap A})) \neq p_2(Z_k | W'(a_{H' \cap A}))$.
But there are sufficient degrees of freedom to satisfy both properties.  In particular, we can choose $p_1(Z_k | W'(a_{H' \cap A}))$ and $p_2(Z_k | W'(a_{H' \cap A}))$ to be distinct one-to-one mappings (since these are 2 by 2 matrices, and almost all such matrices are full column rank) and $c \neq k$ to obtain the above inequality.
%most $2$ by $2$ matrices are full column rank, and thus one to one, and most values of $c,k$ lead to the above inequality.

If $W_k \neq C_k$, the above construction may be trivially extended by letting all variables on the directed path from $C_k$ to $W_k$ be identity functions of their parents.

%As before, it suffices to choose $p_1(Z_{k+1}(a_{H' \cap A}) | Z_{k}(a_{H' \cap A}),W_k) = p_2(Z_{k+1}(a_{H' \cap A}) | Z_{k}(a_{H' \cap A}),W_k)$, so the conditional distributions on the left hand sides are not equal.
%%As before, it suffices to pick the distribution for $Z_{k},Z_{k+1},C_k,W_k$, and variables witnessing that $C_k$ is ancestral of $W_k$ in
%%${\cal G}'_{H'}$ in such a way that
%%$p_1(Z_{k+1}(a_{H' \cap A}) | Z_{k}(a_{H' \cap A})) = p_2(Z_{k+1}(a_{H' \cap A}) | Z_{k}(a_{H' \cap A}))$ is a full column rank matrix
%%with appropriate normalization properties (recall that for binary $Z_k,Z_{k+1}$, this distribution is a 2 by 2 matrix for a fixed $W_k$).
%This establishes the induction.

Assume there exists a recanting district $D$ in ${\cal D}({\cal G}_{W^*})$, where $W^* \equiv \An_{{\cal G}_{V \setminus A}}(W)$.  Further, find the minimal subset $W'$ of $W$ such that the set of all childless vertices in $D$ is in $\An_{{\cal G}_{V \setminus A}}(W')$.  Let $H$ be the smallest set of vertices that contains $W'$, $D$, an element $A_i \in A \cap \Pa_{\cal G}(D)$ with a conflicting treatment assignment, and such that the set of childless vertices in $D$ is in $\An_{{\cal G}_{H}}(W')$.

Consider any edge subgraph of ${\cal G}_H$ such that each vertex has at most one child.
We construct elements $p_1({\mathbb H})$ and $p_2({\mathbb H})$ in the causal model in ${\cal G}_H$ as follows.
In $p_1({\mathbb H})$ each structural equation is a bit parity function of the parents,
and each bidirected arc between $V_i$ and $V_j$ corresponds to a binary latent common parent $\epsilon_{ij}$
where each such latent is involved in precisely two functions.  Moreover $p(\epsilon_{ij})$ is a uniform distribution.
In $p_2({\mathbb H})$ the same is true, except $A_i$ is not involved in the structural equation for any element in $D$.  It has been shown in \cite{shpitser13cogsci} that $p_1(H) = p_2(H)$, but $p_1(W'(\pi, a_i, a'_i)) \neq p_2(W'(\pi, a_i, a'_i))$.

As before, consider a \emph{backdoor path} $\pi$ in ${\cal G}(a)$ from some element $W_i$ in $W'$ to an element $Y_j$ in $Y$, such that $W_i$ is m-connected to $Y$ given $W$, and the edge on the path adjacent to $W_i$ has an arrowhead into $W_i$.
Such a path must exist by construction of $W$.
In addition, consider the smallest subset $W''$ of $W$ such that $W_i$ is m-connected to $Y_j$ given $W''$ in
${\cal G}_{V \setminus A}$.  Pick the smallest set $H'$ containing $H$ such that the above m-connection statement holds in ${\cal G}_{H'}$.
We now extend $p_1({\mathbb H})$ and $p_2({\mathbb H})$ to $p_1({\mathbb H}')$ and $p_2({\mathbb H}')$ to show
$p(Y_j(a_{H' \cap A}) \mid W'(a_{H' \cap A}))$ is not identified.

We have three base cases.  The first case assumes the first node $Z_j$ on $\pi$ not in $H$ is a parent of an element $Z_i$.  In this case, we
let $Z_i$ be the bit parity function of all its parents in ${\cal G}_{H'}$, including $Z_j$ in both $p_1({\mathbb H}')$ and $p_2({\mathbb H}')$.
By reasoning analogous to the hedge case, this implies $p_1(H') = p_2(H')$, but
$p_1(Z_j(a_{H' \cap A}) = 0 \mid W'(a_{H' \cap A}) = 0) < 1$, while $p_2(Z_j(a_{H' \cap A}) = 0 \mid W'(a_{H' \cap A}) = 0) = 1$.

The second case assumes the first node $Z_j$ on $\pi$ not in $H$ is a child of an element $Z_i$ in $H$.  The third case, which we also consider here, assumes $Y \in H$, in which case we let $Y = Z_i$.  If $p(Z_j(\pi, a_i, a'_i) \mid W'(\pi, a_i, a'_i))$ (or $p(Y(\pi, a_i, a'_i) \mid W'(\pi, a_i, a'_i))$) is not identified, we are done.  Otherwise, we assume $p(Z_j(\pi, a_i, a'_i) \mid W'(\pi, a_i, a'_i))$ is identified.  Consider the edge subgraph ${\cal G}'_{H'}$ of ${\cal G}_{H'}$ that lacks the outgoing directed edges from $Z_i$ within $H$.

If $Z_i$ is not in $D$, by reasoning analogous to reasoning in the hedge case, ${\cal G}_{H'}$ contains the recanting district structure for $p(Z_j(\pi, a, a'), W''(\pi, a, a'))$, where $W'' = W' \setminus \{ W_i \}$ and $W_i$ is $W' \cap \De_{{\cal G}_{H'}}(Z_j)$.  If $p(W''(\pi, a, a'))$ is identified, we are done, since we established the base case where $p(Z_j(\pi, a, a') \mid W''(\pi, a, a'))$ is not identified.  If $p(W''(\pi, a, a'))$ is not identified, note that $W''$ is a strictly smaller set then $W'$, and we restart the base case argument, finding either a hedge or a recanting district for this smaller set, constructing a new set $H$, and a new backdoor path to an element in $Y$.  Since the new subset of $W$ is strictly smaller, we can only do this a finite number of times before encountering another base case.

If $Z_i$ is in $D$, then the resulting graph ${\cal G}'_{H'}$ contains a recanting district structure for $p(Z_j(\pi, a, a'), W'(\pi, a, a'))$ with the set of childless vertices of the previous district and also $Z_i$ (since it is now childless in $H$).
Given the recanting district construction, $p_1(Z_j(\pi, a, a') = 0 \mid W'(\pi, a, a') = 0) < 1$, while
$p_2(Z_j(\pi, a, a') = 0 \mid W'(\pi, a, a') = 0) = 1$, and we are done.

Since we now established bases for the induction for the recanting district case, we can apply the inductive argument for the hedge case to conclude $p(Y(\pi, a_i, a_i') \mid W'(\pi, a_i, a_i'))$ is not identified, as above.  Having established that $p(Y(\pi, a_i, a_i') \mid W'(\pi, a_i, a_i'))$ is not identified in ${\cal G}_{H'}$ or ${\cal G}'_{H'}$, it is trivial to extend $p_1({\mathbb H}')$ and $p_2({\mathbb H}')$ to $p_1({\mathbb V})$ and $p_2({\mathbb V})$ for ${\cal G}(V)$.

Finally, our conclusion is established for ${\cal G}^e(V, A^{\Ch})$ and $p(Y(a^{\pi}) \mid W(a^{\pi}))$ by Proposition \ref{prop:equiv}.
\end{prf}

\section*{A Weaker Causal Model}
%\label{sec:weaker}

We phrased all our discussion in terms of the functional causal model, defined by the restriction (3). %(\ref{eqn:mwm}).
A weaker causal model called the \emph{finest fully randomized causally interpretable structured tree graph} (FFRCISTG) suffices for many causal inference tasks. This model asserts that the variables,
\begin{align}
\left\{
V_i({\pa_i}) \mid i \in \{ 1, \ldots, k \}
\right\},
\label{eqn:swm}
\end{align}
are mutually independent for every $v \in {\mathfrak X}_V$, where $\pa_i$ is the subset of $v$ associated with $\Pa_i$.  
Note that the set of independences
asserted by (\ref{eqn:swm}) 
is a subset of the set of independences asserted by (3).  
In particular, (\ref{eqn:swm})
only asserts independences among a set of potential outcomes associated with a globally consistent intervention operation, while
(3) may allow independences among potential outcomes with inconsistent interventions.  For example, a model defined by (3) may assert that $Y(a,m) \ci M(a')$, while (\ref{eqn:swm}) 
never asserts such an independence if $a \neq a'$.

Since the SWIG global Markov property only asserts independences on random variables associated with a globally consistent intervention operation, it is implied not only by (3) but also the weaker model represented by (\ref{eqn:swm}) \cite{thomas13swig}.  Potential outcomes like $p(Y(a,M(a')))$ that arise in mediation analysis are not identified under (\ref{eqn:swm}), but are sometimes identified under (3); see \cite{shpitser15hierarchy} for details.  Note, however, that our rephrasing of edge-consistent counterfactuals $p(V_i(\pi, a, a'))$ in the causal model for ${\cal G}(V)$ in terms of an intervention $p(V_i(a^{\pi}))$ in the extended causal model for ${\cal G}^e(V \cup A^{\Ch})$ leads to an identification theory for which model (\ref{eqn:swm}) 
for the variables in $\mathbb{V}^e$ is sufficient. The reason that counterfactuals $p(V_i(\pi, a, a'))$ requiring the stronger set of assumptions (3)
may be rephrased as counterfactuals $p(V_i(a^{\pi}))$ only requiring the weaker set of assumptions (\ref{eqn:swm}) has to do with the specific way in which ${\cal G}^e$ was constructed.  Specifically, ${\cal G}^e$ implicitly imposed strong restrictions on the associated FFRCISTG, having to do with deterministic relationships between $A_i$ and $A^j_i$ as well as absences of edges between any element $A_i^j$ in $A^{\Ch}$ and any element in $\Ch^{\cal G}_i$ other than $V_j$.  Had these edges not been absent in ${\cal G}^e$, identification would no longer be possible.  In some sense, ${\cal G}^e$ is the graph corresponding to the ``weakest'' FFRCISTG that encodes assumptions associated with the functional model on ${\cal G}$.  These assumptions may be viewed informally as stating that a treatment variable $A_i$ in $A$ may be decomposed into components that only influence particular children (immediate effects) of $A_i$, and no other children of $A_i$.

\small
\bibliographystyle{plain}
\bibliography{references}